\documentclass[11pt,manuscript]{aastex}
\usepackage{epsf}

\usepackage{ifthen}
\newcounter{emulate}
\ifthenelse{\value{emulate}=1}{
  \usepackage{emulateapj5,apjfonts}
  \usepackage{onecolfloat}
}{}

\received{24 April 2000}
\accepted{3 August 2000}

\shortauthors{de Jong \& Lacey}
\shorttitle{The bivariate space density function of spiral galaxies}

\newcommand{\degree}{\arcdeg}
\newcommand{\magarc}{mag arcsec$^{-2}$}

\newcommand{\muo}{\relax \ifmmode \mu_{\rm 0}\else $\mu_{\rm 0}$\fi}
\newcommand{\ncsbd}{\relax \ifmmode \mu_{\rm 0}\else $\mu_{\rm 0}$\fi}
\newcommand{\mue}{\relax \ifmmode \mu_{\rm e}\else $\mu_{\rm e}$\fi}
\newcommand{\muave}{\relax \ifmmode \langle\mu\rangle_{\rm e}\else $\langle\mu\rangle_{\rm e}$\fi}
\newcommand{\magtot}{\relax \ifmmode M_I\else $M_I$\fi}
\newcommand{\magd}{\relax \ifmmode M_{I,D}\else $M_{I,D}$\fi}
\newcommand{\re}{\relax \ifmmode r_{\rm e}\else $r_{\rm e}$\fi}
\newcommand{\red}{\relax \ifmmode r_{\rm e,D}\else $r_{\rm e,D}$\fi}
\newcommand{\res}{\relax \ifmmode r_{\rm e*}\else $r_{\rm e*}$\fi}
\newcommand{\h}{\relax \ifmmode h\else $h$\fi}

\newcommand{\muoi}{\relax \ifmmode \mu^i_{\rm 0}\else $\mu^i_{\rm 0}$\fi}
\newcommand{\ncsbdi}{\relax \ifmmode \mu^i_{\rm 0}\else $\mu^i_{\rm 0}$\fi}
\newcommand{\muei}{\relax \ifmmode \mu^i_{\rm e}\else $\mu^i_{\rm e}$\fi}
\newcommand{\muavei}{\relax \ifmmode \langle\mu\rangle^i_{\rm e}\else $\langle\mu\rangle^i_{\rm e}$\fi}
\newcommand{\magtoti}{\relax \ifmmode M^i_I\else $M^i_I$\fi}
\newcommand{\magdi}{\relax \ifmmode M^i_{I,D}\else $M^i_{I,D}$\fi}
\newcommand{\rei}{\relax \ifmmode r^i_{\rm e}\else $r^i_{\rm e}$\fi}
\newcommand{\redi}{\relax \ifmmode r^i_{\rm e,D}\else $r^i_{\rm e,D}$\fi}

\newcommand{\vmax}{\relax \ifmmode V_{\rm max}\else $V_{\rm max}$\fi}
\newcommand{\rqua}{$R^\frac{1}{4}$}
\newcommand{\zq}{(1+z)$^4$}

\begin{document}

\submitted{Accepted for publication in the Astrophysical Journal}

\newcommand{\head}{

\title{The local space density of S\lowercase{b}-S\lowercase{dm} galaxies as
function of their scalesize, surface brightness and luminosity}

\author{Roelof S.\ de Jong\altaffilmark{1}}
\affil{Steward Observatory, 933 N.\ Cherry Ave., Tucson, AZ 85721, USA}
\affil{University of Durham, Department of Physics, South Road, Durham DH1 3LE, UK}
\affil{email: {\tt rdejong@as.arizona.edu}}
\email{rdejong@as.arizona.edu}
\and
\author{Cedric Lacey}
\affil{TAC, Juliane Maries Vej 30, DK-2100 Copenhagen O, Denmark}
\affil{University of Durham, Department of Physics, South Road, Durham DH1 3LE, UK}
\affil{SISSA, via Beirut, 2-4, 34014 Trieste, Italy}
\affil{email: {\tt lacey@sissa.it}}
\email{lacey@sissa.it}
\altaffiltext{1}{Hubble fellow}

\begin{abstract}
We investigate the dependence of the local space density of spiral
galaxies on luminosity, scalesize and surface brightness. We derive
bivariate space density distributions in these quantities from a
sample of about 1000 Sb-Sdm spiral galaxies, corrected for selection
effects in luminosity and surface brightness. The structural
parameters of the galaxies were corrected for internal extinction
using a description depending on galaxy surface brightness. We find
that the bivariate space density distribution of spiral galaxies in
the (luminosity, scalesize)-plane is well described by a Schechter
luminosity function in the luminosity dimension and a log-normal scale
size distribution at a given luminosity. This parameterization of the
scalesize distribution was motivated by a simple model for the
formation of disks within dark matter halos, with halos acquiring
their angular momenta through tidal torques from neighboring objects,
and the disk specific angular momentum being proportional to that of
the parent halo. However, the fractional width of the scalesize
distribution at a given luminosity is narrower than what one would
expect from using the distribution of angular momenta of halos
measured in N-body simulations of hierarchical clustering. We present
several possible explanations for the narrowness of the observed
distribution. Using our bivariate distribution, we find that
determinations of the local luminosity function of spiral galaxies should
not be strongly affected by the bias against low surface brightness
galaxies, even when the galaxies are selected from photographic
plates. This may not be true for studies at high redshift, where \zq\
surface brightness dimming would cause a significant selection bias
against lower surface brightness galaxies, if the galaxy population
did not evolve with redshift.
\end{abstract}

\keywords{%
galaxies: formation --
galaxies: fundamental parameters --
galaxies: luminosity function --
galaxies: spiral --
galaxies: statistics --
galaxies: structure
}

} 

\ifthenelse{\value{emulate}=1}{
\twocolumn[\head]
}
{
\head
}

\section{INTRODUCTION}

In the last few decades, many papers have been devoted to the
measurement of the luminosity function (LF) of galaxies, of their
distribution of central surface brightnesses and, to a lesser extent,
of their distribution of scalesizes. The observational determinations
of these three types of distribution cannot in practice be separated,
because of the limitations of the surveys on which the investigations
are based.  Any galaxy LF is only valid to the surface brightness
limit of the survey from which it is derived, while any distribution
of surface brightnesses is valid only over some range in luminosity or
scalesize, depending on the survey limits in apparent magnitude and/or
angular size. In this paper we address this problem directly, by
investigating the bivariate distribution functions of spiral galaxies
in combinations of luminosity, surface brightness and scale
size. Knowledge of any two of these quantities then suffices to
determine the third.

Bivariate distribution functions have two important applications.
First of all, bivariate distribution functions are the only proper way
to compare samples with different selection criteria, especially when
comparing samples at different redshifts.  For instance, comparing LFs
determined from samples with similar magnitude limits but different
lower surface brightness limits will result in discrepancies in the
magnitude range where the contribution from low surface brightness
galaxies is significant.  Secondly, bivariate distribution functions
provide excellent tests for galaxy formation and evolution theories.
Any complete galaxy formation theory should be able to explain the
distribution functions of galaxy structural parameters.  Obviously,
the 2-dimensional (2D) distribution functions provide more constraints
on formation theories than the separate 1D distributions of surface
brightness, scalesize and luminosity obtained by integrating over the
other quantity in the bivariate distribution.

As already mentioned, every optically-selected galaxy sample always
has limits in surface brightness in addition to its limits in apparent
luminosity and/or angular diameter. The detection volume (or
visibility) for a particular type of galaxy in a such a survey is then
at least a two-parameter function, e.g.\ of luminosity and scalesize, and
depends strongly on these parameters, resulting in strong biases
against low surface brightness (LSB) and small scalesize galaxies
(Disney \& Phillipps~\citeyear{DisPhi83}; Allen \& Shu
\citeyear{AllShu79}; McGaugh, Bothun \&
Schombert~\citeyear{McG95}). Since the determination of the space
density of galaxies from a survey depends on knowing the detection
volumes, the only complete description of the galaxy space density
which can be obtained observationally is a bivariate distribution
function which includes two of the three parameters of surface
brightness, scalesize and luminosity. To study the bivariate
distribution of field spiral galaxies, it is straightforward to show
that one has to obtain surface photometry {\em and} distances of at
least 500-1000 galaxies, in order to avoid problems with small number
statistics near the selection boundaries (de Jong \& Lacey
\citeyear{deJLac99}).

Because of the large number of galaxies needed with both redshifts and good
surface photometry, determinations of bivariate distribution functions of
spiral galaxies as functions of structural parameters have been
relatively rare. Some notable exceptions are Phillipps \& Disney
(\citeyear{PhiDis86}), who presented a (magnitude, surface
brightness)-distribution of Virgo spiral galaxies using RC2 data, van
der Kruit (\citeyear{vdK87}), who used a diameter-limited sample of 51
galaxies to construct a crude (surface brightness, scale
length)-distribution, and Sodr\`e \& Lahav (\citeyear{SodLah93}) who
created a (magnitude, diameter)-diagram from the ESO-LV catalog. More
recently Lilly et al.~(\citeyear{Lil98}) used the CFRS redshift survey
to derive the bivariate function in the (magnitude, scalesize)-plane,
and made a first attempt at studying its redshift evolution. Finally,
Driver (\citeyear{Dri99}) used a volume-limited selection of galaxies
in the Hubble Deep Field (Williams et al.~\citeyear{Wil96}) to probe
the really low surface brightness regime of the bivariate distribution
function.  The results of nearly all of these studies suffered from
small number statistics, and very few firm physical conclusions could
be drawn.

Theoretical predictions for the sizes of galaxy disks in the
hierarchical clustering picture of galaxy formation began with the
classic paper by Fall \& Efstathiou (\citeyear{FalEfs80}). They
considered the formation of a disk by the collapse of gas within a
gravitationally dominant dark matter (DM) halo. They showed how the radius
of the disk is related to that of the halo, on the assumption that the
gas starts off with the same specific angular momentum as the dark
matter, and conserves its angular momentum during the collapse. Thus
in this picture, the disk radius depends on the amount of angular
momentum which the halo acquires prior to collapse through the action
of tidal torques from neighboring objects. This model naturally leads
to typical disk sizes similar to those observed for bright spiral
galaxies.  Many authors have subsequently made calculations of disk
sizes within the same basic framework (see e.g.\ van der Kruit
\citeyear{vdK87}; Mo et al.~\citeyear{Mo98}; van den Bosch
\citeyear{vdB98}), and Dalcanton et al.~\citeyear{Dal97b} combined this
model with a Schechter luminosity function for galaxies to predict the
bivariate distribution of surface brightness and scale length for
disks. We will parameterize our observed bivariate distribution
function for disks in a way that is motivated by this same simple
model.

More recently, predictions for galaxy properties in hierarchical
clustering models have been developed much further using the technique
of semi-analytic modeling (Cole et al.~\citeyear{Col94},
\citeyear{Col00}; Kauffmann, White \& Guiderdoni~\citeyear{Kau93},
Somerville \& Primack~\citeyear{Som99}). The semi-analytic models
include much more of the physics of galaxy formation, including the
merging histories of DM halos, gas cooling and collapse within halos,
star formation from cold gas, feedback from supernovae, and the
luminosity evolution of stellar populations.  In this paper we will
compare the observed bivariate distribution function with the most
recent semi-analytic model predictions from Cole et
al.~(\citeyear{Col00}).

This paper is organized as follows. In \S\ref{viscor} we describe how
one can correct a sample of objects for distance dependent selection
effects. In \S\ref{sample} we describe the sample we have used for
this investigation and how we determine physical quantities from
the observations. In \S\ref{bivar} we determine the bivariate
distributions of space density and luminosity density for the local
universe. We propose a model for the bivariate distribution functions
based on the hierarchical galaxy formation scenario and fit this model to
the data in \S\ref{func}. Finally, we discuss the results in
\S\ref{discuss} and summarize our conclusions in \S\ref{concl}.

\section{VISIBILITY CORRECTION}
\label{viscor}

The use of selection criteria to define a sample of objects often
introduces selection biases, even in so called ``complete samples'',
i.e.\ samples that are complete according to their selection
criteria. Malmquist (\citeyear{Mal20}) was one of the first to
quantify the bias in the determination of the average {\it absolute}
magnitude of a stellar sample due to the real spread in luminosity
combined with the distance-dependent selection limit that results from
applying a cut-off in {\it apparent} magnitude. The uncertainty in the
measurement of the selection magnitude introduces another bias near
the selection limit, which can be described in the same way as
Malmquist's original bias if the uncertainties have a Gaussian error
distribution. Both of these effects (which are mathematically similar
in case of Gaussian luminosity and error distributions, but have
completely different origins) have been called Malmquist bias by
different authors. To make matters even more confusing, the biases in
distance measurements resulting from the use of samples suffering from
these effects have also been called Malmquist bias.

In this section we describe how to correct a sample for
distance-dependent biases and for biases resulting from uncertainties
in the selection parameters. We pay particular attention to the case
where the sample has been selected on angular diameters.

\subsection{Volume Correction}

Our aim is to determine the average space density of galaxies with
certain properties in the local universe.  Most field galaxy samples
are not based on distance- or volume-limited surveys, but are limited
by some quantity more readily available observationally, like apparent
magnitude or angular diameter.  Not all galaxies have the same
luminosity or physical diameter, and therefore they can be seen to
different distances before dropping out due to the selection limits.
The volume within which a galaxy can be seen and will be included in
the sample (\vmax) goes as the distance limit cubed, which results in
galaxy samples being dominated by intrinsically bright and/or large
galaxies, because these have the largest visibility volume (Disney \&
Phillipps \citeyear{DisPhi83}; McGaugh et al.~\citeyear{McG95}).

In this paper we use one of the simplest methods available for
correcting for selection effects, the \vmax\ correction method
(Schmidt \citeyear{Sch68}).  Each galaxy is given a weight equal to
the inverse of its maximum visibility volume set by the selection
limits (a formal derivation can be found in Felten
(\citeyear{Fel76})).  For a low redshift sample with upper ($D_{\rm
max}$) and lower ($D_{\rm min}$) limits on the major axis angular diameter,
this leads to
\begin{equation} 
\vmax = \Omega_f\frac{4\pi}{3}d^3 \left[ \left(
\frac{D_{\rm maj}}{D_{\rm max}}\right)^3 -\left(\frac{D_{\rm
maj}}{D_{\rm min}}\right)^3\right]
\label{vmax}
\end{equation} 
with $\Omega_f$ the fraction of the sky used to select the galaxies,
$d$ the distance to the galaxy, and $D_{\rm maj}$ the major axis
angular diameter of the galaxy.  Other limits, like redshift or
magnitude limits, that would limit \vmax\ can trivially be taken into
account as well.  For higher redshift samples we have to take
cosmological corrections into account. We define the bivariate density
distribution $\phi(x,y)$ in parameters $x$ and $y$ such that
$\phi(x,y)\,dx\,dy$ is the number density of galaxies in the interval
$(x,y),(x+dx,y+dy)$. For a sample of $N$ galaxies which is complete to
within the selection limits, we can now define an estimator of this
quantity as follows:
\begin{equation} 
\phi(x,y) \approx \frac{1}{\Delta
 x\Delta y} \sum_{i}^N \frac{\delta^i}{\vmax^i},
\label{bivarest}
\end{equation} 
where $i$ is summed over all galaxies, and $\delta^i=1$ if the
($x_i$,$y_i$) parameters of galaxy $i$ are in the bin range ($x\pm
\Delta x/2,y\pm \Delta y/2$), and 0 otherwise.

The \vmax\ correction method assumes a uniform distribution of
galaxies in space, and is not unbiased against density fluctuations.
To give unbiased results, objects with the smallest \vmax\ in the
sample should be visible on scales larger than the largest scale
structure.  Currently, such samples do not exist.  Other methods exist
that take density fluctuations into account (for reviews see e.g.\
Efstathiou et al.~\citeyear{Efs88}; Willmer \citeyear{Wil97}).  These
methods assume a direct relation between the distribution parameter
and the selection parameter.  This is not the case in the current
investigation (selection on $B$-band diameters versus distributions of
$I$-band magnitudes, surface brightnesses and scalesizes).

The \vmax\ corrections of equations (\ref{vmax}) and (\ref{bivarest})
are valid only if similar galaxies have their angular diameters
measured at the same physical diameter, independent of distance (see
the discussion in de Jong \citeyear{deJ3}). It is {\em not} important
if a particular class of galaxies has their diameters measured at an
intrinsically larger physical diameter compared to other classes (for
instance, at a lower surface brightness). This class of galaxies will
be over-represented in the sample, but on average will have a larger
distance, so that the effects exactly cancel out in the
estimator~(\ref{bivarest}), as they are designed to do. In a similar
fashion to de Jong \& van der Kruit (\citeyear{deJ1}), we determined
that the ratio of eye-estimated to isophotal diameters was independent
of diameter, and we therefore conclude that most likely the diameters
of all similar galaxies were measured at the same physical (linear)
diameter.

The uncertainty in the $\phi(x,y)$ estimator of
equation\,(\ref{bivarest}) is in general dominated by Poisson
statistics: what is the uncertainty on the mean number of galaxies in
a bin, if $N$ are detected?  It is easy to show that at least 500
galaxies with accurate photometry and distances are needed to
determine a bivariate distribution function of structural parameters
(de Jong \& Lacey \citeyear{deJLac99}).  Only if we have many galaxies
in a bin is the error in $\phi(x,y)$ no longer dominated by Poisson
statistics, but becomes dominated by the uncertainty in \vmax.  The
uncertainty in \vmax\ in equation\,(\ref{vmax}) arises from galaxy
distance uncertainties and diameter uncertainties.  The distance
uncertainty of each galaxy ($\sigma^i_d$) contributes a component
$\sum_{i}^N(\frac{\sigma^i_d}{d^i} \frac{3\delta^i}{\vmax^i})^2$ to the variance
in the determination of $\phi(x,y)$.  The diameter uncertainties
($\sigma^i_{D_{\rm maj}}$) add a
$\sum_{i}^N(\frac{\sigma^i_{D_{\rm maj}}}{D^i_{\rm maj}} \frac{3\delta^i}{\vmax^i})^2$
to the variance, but are on top of that directly related to the
selection of the sample, and are further discussed in the next section.

\subsection{Selection Uncertainty Correction}

The parameters used to select the galaxy sample can only be determined
with finite accuracy.  The selection parameters have to be distance
dependent for \vmax\ corrections to be used, leading to what often is
called the Malmquist edge bias.  Assuming a symmetric error
distribution on the selection parameters (e.g.\ diameter or magnitude),
objects at the selection limit have an equal probability of being
scattered into the sample as being scattered out of the sample.
Because there are more small and faint than large and bright objects
on the sky (due to the effect described in the previous paragraph),
on average more objects are scattered into the sample than than out of
the sample, and we will overestimate the number of objects
in our search volume.

Once we have determined the probability distribution of the error in
the selection parameters ($P(x)$), we can correct the \vmax\ method
for this edge bias.  We might try to correct for the bias by taking
the average 1/\vmax\ weighted with the error distribution of the
selection parameters within the selection limits.  Unfortunately, this
procedure would result in an overcorrection.  An object at the
selection limit would count for only half (with the other half being
outside the selection limits assuming a symmetrical error
distribution), but an object just outside the selection limit with a
large fraction of its probability function within the limits would not
be included at all.  To remedy this effect, we take a virtual
selection limit $t\sigma$ away from the original selection limit and
now include the objects between the virtual and original selection
limits with appropriate (low) weight.

\ifthenelse{\value{emulate}=1}{
\begin{figure}[tb]
\epsfxsize=\linewidth \epsfbox[30 150 570 700]{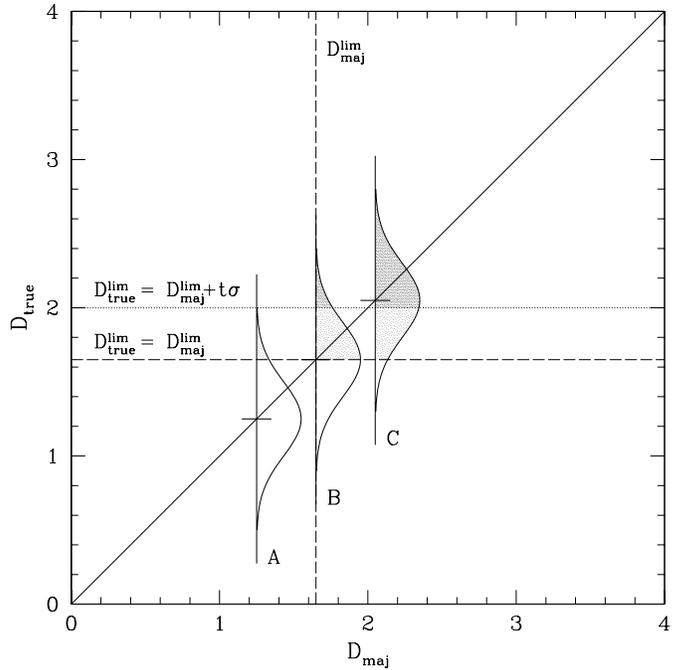}
\figcaption[f1.ps]{Observed diameters versus the probability distributions
of true diameters for three galaxies. The solid diagonal line is the
line of equality, the dashed and dotted lines indicate different
minimum diameter selection limits in observed and true angular
diameters. For detailed explanation see text.
\label{malmquist}
}
\end{figure}
}{}

We demonstrate this for the case of diameter selection with the use of
Fig.\,\ref{malmquist}, where we plot three galaxies of different
observed diameters $D_{\rm maj}$. Galaxy A has an observed diameter
just below the selection limit $D_{\rm min}$ (indicated by the
vertical dashed line), galaxy B is at the selection limit and galaxy C
has an observed diameter slightly larger than the selection
limit. Each galaxy has an associated probability distribution of true
diameter $D_{\rm true}$, indicated by the Gaussian distributions. We
can calculate a corrected 1/\vmax\ for each galaxy using
equation\,(\ref{vmax}), averaged over the range of true diameters for
each galaxy, weighted appropriately for the true diameter
probability. If we take the true diameter cut-off the same as the
observed diameter cut-off (indicated by the horizontal dashed line),
than galaxy B is only counted for half a galaxy, galaxy C is counted
almost completely, and galaxy A is counted for a small but significant
fraction (dark and light shaded regions). We now have the situation
where galaxy A should have been included, because a significant part
of its true diameter distribution is larger than the true diameter
selection limit, but the galaxy is in fact not included in the sample
at all because its observed diameter is below the selection limit. This
attempt to correct for the edge bias is therefore wrong, as we are not
counting galaxies that should have been included. But by shifting the
virtual true diameter selection limit upwards (indicated by the dotted
line) and calculating the \vmax\ values from equation\,(\ref{vmax})
with this shifted diameter limit, we only have to weight the galaxies
for the dark shaded regions. Galaxy A has now a negligible fraction of
true diameters above the true diameter selection limit, which is good
because it was not in the sample to begin with. Other galaxies just
above the selection limit get little weight, but have the appropriate
corrected 1/\vmax\ values.

In our example of a complete sample selected with upper and lower
angular diameter cut-offs $D_{\rm max}$ and $D_{\rm min}$, we get for the
corrected 1/\vmax to use in equation\,(\ref{bivarest})
 \begin{equation}
  1/\vmax^{\rm cor} = 
  \frac{\int_{D_{\rm min}+t\sigma}^{D_{\rm max}-t\sigma}
       P(D|D_{\rm maj})/\vmax(D)\, dD
      }
      {\int_{-\infty}^{\infty} P(D|D_{\rm maj})\, dD
      },
  \label{unccor}
 \end{equation}
where $P(D|D_{\rm maj})$ denotes the probability of the true
angular diameter of a galaxy being $D$ at a given observed angular
diameter $D_{\rm 
maj}$, and \vmax(D) is to be evaluated using equation\,(\ref{vmax}) with
$D_{\rm min}$ replaced by $D_{\rm min}+t\sigma$ and $D_{\rm
max}$ replaced by $D_{\rm
max}-t\sigma$.  We should try to make $t\sigma$ as large as possible,
so that $P(D\pm t\sigma|D)$ is small, and the probability of a galaxy
apparently being outside the selection limits but in reality belonging
inside is small. Unfortunately, we cannot make $t\sigma$ too large, as
then very few galaxies will remain with significant weights.

\section{SAMPLE SELECTION AND DATA}
\label{sample}

We have used the galaxy sample described by Matthewson, Ford and
Buchhorn (\citeyear{MFB}) and Matthewson \& Ford (\citeyear{MatFor96},
MFB sample hereafter) as the starting point for our sample selection.
The MFB data were obtained mainly to study peculiar motions using the
Tully-Fisher (\citeyear{TulFis77}) relation.  The MFB sample is nearly
ideal for the kind of study we want to perform.  With more than a
thousand field galaxies it is large enough not to run immediately into
small number statistics near the low surface brightness and/or small
scalesize selection borders.  Matthewson et al.\ collected for most
objects the CCD surface photometry and redshifts required for our
statistical study.  The main drawback of the sample is its selection,
as the sample was defined as a subsample of the ESO-Uppsala Catalog
of Galaxies (Lauberts \citeyear{Lau82}), which is a catalog selected by eye
from photographic plates.  Unfortunately, nothing better exists at the
moment, and it remains to be seen whether automated surveys like
Sloan, DENIS and 2MASS will go deep enough to detect LSB galaxies.
These surveys should however discover and quantify the number of
galaxies with small scalesizes.

The MFB sample is not entirely complete, as some selected galaxies had
to be excluded due to too bright foreground stars, too disturbed
morphologies to obtain reliable surface brightness profiles or
inability to obtain redshifts.  As incompleteness is an issue in our
analysis, we went back to ESO-Uppsala catalog and reselected galaxies
using selection criteria close to the MFB sample criteria.  Our
criteria are: ESO-Uppsala diameter $1.65\arcmin$\,$\le$\,$D_{\rm
maj}$\,$\le$\,$5.05\arcmin$, galactic latitude $|b|> 11\degree$,
morphological type $3$\,$\le$T$\le$\,$8$ and minor-over-major axis
ratio $0.1736$ $<$\,$D_{\rm min}/D_{\rm maj}$\,$<$\,$0.776$.  This
last criterion is different from MFB, excluding the edge-on galaxies
for which extinction corrections are large and uncertain.  These
selection criteria resulted in a sample of 1007 galaxies, with a
subsample of 818 galaxies (81.2\%) for which we have both MFB surface
photometry and redshifts (some redshifts were obtained from the NED
and LEDA databases).

A $V/\vmax$-test (Schmidt \citeyear{Sch68}; see also de Jong
\citeyear{deJ3}) corrected for Malmquist edge bias showed that the
sample has an average $V/\vmax$ of 0.507$\pm$0.010 and is therefore
statistically complete.  A slight incompleteness for high surface
brightness galaxies ($\langle V/\vmax\rangle=0.454\pm0.021$ for
galaxies with \hbox{$\muo$\,$<$\,$19$\,$I$-\magarc}) was detected.
This means we have either too many high surface brightness galaxies
nearby or too few at large distance.  We could find no obvious reason
why this might be the case. For the lower surface brightness bins the
$V/\vmax$ indicated statistical completeness.

Accurate distances are essential to calculate the \vmax\ corrections.
Applying blind Hubble flow distances would introduce large errors for
many of the smallest, nearby galaxies.  Luckily, because the MFB
sample data were obtained to measure peculiar motions, many of our
galaxies have Tully-Fisher distances (Tully \&
Fisher~\citeyear{TulFis77}).  For the 706 galaxies in
our sample also included in the Mark III catalog (Willick et
al.~\citeyear{MarkIII97}) we used group velocities for groups with
recession velocities larger than 2000\,km/s, otherwise the Mark III
Malmquist bias corrected velocities.  For galaxies not included in the
Mark III catalog, we used their Heliocentric velocity corrected to the
Local Group velocity according to the precepts of Karachentsev \&
Makarov (\citeyear{KarMak96}).  All these velocities were converted to
distances using a Hubble constant of 65 km\,s$^{-1}$ Mpc$^{-1}$. When
calculating the \vmax\ corrections, we assume a 15\% distance error
for the galaxies with Mark III velocities and a 250 km/s peculiar
velocity uncertainty for the remaining galaxies (1 sigma
uncertainties).

Twenty percent of the galaxies have velocities of less than 2000\,km/s
and about another 20\% have velocities exceeding 5000\,km/s.  For the
brightest galaxies we therefore sample large enough scales not to be
influenced by large scale density fluctuations, but for smaller galaxies
this may not be the case.  However, because $V/\vmax$-tests indicated
completeness and homogeneity of the sample independent of surface
brightness and scalesize, we are not overly concerned by this. 

We calculated the characteristic global structural parameters of the
galaxies from the radial $I$-band luminosity profiles.  MFB calculated
luminosity profiles by determining the average surface brightness on
elliptical annuli, which had been fitted to the galaxy isophotes.  The
total luminosity (\magtot) of the galaxies was calculated by
extrapolating the last few measured points of the profiles to infinity
with an exponential luminosity profile.  This luminosity was used to
calculate the effective (half total enclosed light) radius (\re).  The
average surface brightness within the effective radius --which we will
call effective surface brightness (\muave) hereafter-- was calculated
using $\muave = \magtot - 5\log(\re) - 2.5\log(2\pi)$.  

In addition to the structural parameters for the galaxy as a whole, we
also use in this paper the structural parameters for the disk alone.
We decomposed the 1D luminosity profiles into bulge and disk
contributions, using exponential light profiles for both disk and
bulge (see de Jong \citeyear{deJ2}).  This yielded the structural
parameters disk magnitude (\magd), disk central surface brightness
(\muo) and disk effective radius (\red), which equals 1.679 times the
disk e-folding scale length.  In agreement with de Jong
(\citeyear{deJ2}), the 1D disk parameters showed good agreement with
the disk parameters determined by Byun (\citeyear{Byu92}), who used a
2D fitting method and an \rqua\ instead of exponential profile for
the bulge.

The Galactic foreground extinction corrections were calculated
according to the precepts of Schlegel, Finkbeiner \&
Davis~(\citeyear{Sch98}).  The proper internal extinction correction
for disk galaxies is still heavily in debate.  Many different
corrections have been proposed, resulting from a large variety of
methods and galaxy samples.  Here we use the method of Byun
(\citeyear{Byu92}), also described in detail by Giovanelli et
al.~(\citeyear{Gio95}), to correct quantities to face-on values.
Using this method, the parameter for which the extinction correction
has to be determined is first fitted against the inclination corrected
maximum rotation velocity of the disk ($V_{\rm rot}$).  The residuals on this
fit are next fitted against $\log(D_{\rm min}/D_{\rm maj}$) to
empirically determine the effect of extinction as a function of
inclination relative to face-on. The extra step of fitting to the
residuals of the $V_{\rm rot}$ relation reduces the distance dependent
selection effects as function of inclination.

In contrast to Giovanelli et al.~(\citeyear{Gio95}) and Tully et
al.~(\citeyear{Tul98}), we divide the extinction measurements into
several surface brightness bins instead of absolute
magnitude bins, as we expect the amount of extinction to be more related to
surface brightness than luminosity.  If the amount of dust at a given
radius in the galaxy is in some way proportional to the amount of
stars at that radius (i.e.\ local surface brightness), then for a
disk-like configuration the relative extinction as function of
inclination will be determined by surface brightness, independent of
scalesize and hence magnitude. Even so, because the magnitudes and surface
brightnesses of galaxies are to some extend correlated, one will also
see a trend between magnitude and extinction. The equations used for
the extinction
corrections are listed in Table~\ref{exttab}.  We find that the low
surface brightness galaxies in our sample behave as nearly transparent
disks, while high surface brightness disks behave as having optical
depth larger than one near the center. 

\begin{table}[tb]
\caption{Internal extinction corrections}
{\small
\begin{tabular}{rr}
\tableline
\tableline
corr. par.& \multicolumn{1}{c}{equation used}\\
\tableline
\muavei      &$\muave    \,+\,(0.180(\muave-24)-0.030)\log(D_{\rm maj}/D_{\rm min})$\\
\ncsbdi      &$\ncsbd    \,+\,(0.613(\muavei-24)+2.862)\log(D_{\rm maj}/D_{\rm min})$\\ 
$\log(\rei)$ &$\log(\re) \,-\,(0.039(\muavei-24)+0.083)\log(D_{\rm maj}/D_{\rm min})$\\
$\log(\redi)$&$\log(\red)\,+\,(0.019(\muavei-24)+0.036)\log(D_{\rm maj}/D_{\rm min})$\\
\magtoti     &$\magtot   \,+\,(0.197(\muavei-24)-0.058)\log(D_{\rm maj}/D_{\rm min})$\\
\magdi       &$\magd     \,+\,(0.295(\muavei-24)-0.488)\log(D_{\rm maj}/D_{\rm min})$\\
\tableline
\end{tabular}
\label{exttab}
}
\end{table}

\ifthenelse{\value{emulate}=1}{
\begin{figure}[tb]
\epsfxsize=\linewidth \epsfbox[45 172 528 539]{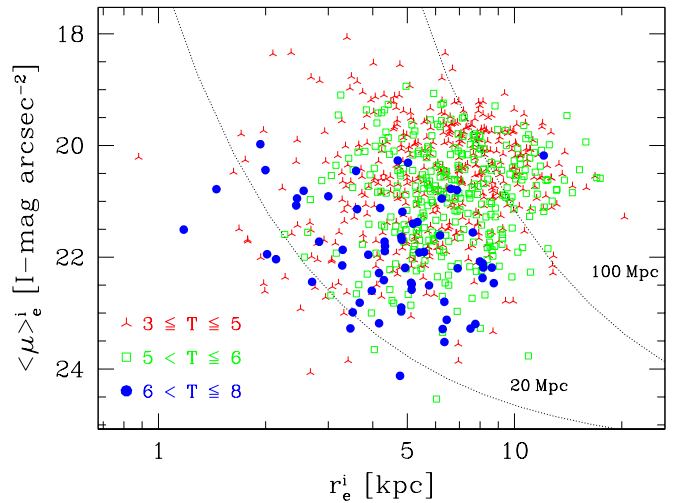}
\figcaption[f2.ps]{Observed distribution of effective surface brightness
versus effective radius, both corrected for extinction to face-on
values. The dotted lines show the maximum indicated distances to which
face-on exponential disk galaxies can be observed given the selection
criteria of our sample. Different symbols are used to denote the
indicate ranges of ESO-Uppsala morphological T-type.
\label{uncordis_re_muave}
}
\end{figure}
}{}

Figure\,\ref{uncordis_re_muave} shows the distribution of the galaxies
as a function of the extinction-corrected values \muavei\ and \rei.
The dotted lines illustrate the selection biases for this sample.  The
sample should be complete to the indicated distances, for galaxies
above and to the right of the lines, if we assume purely face-on
exponential disks.  The lines were calculated assuming the average
surface brightness at $D_{\rm maj}$ is 24.83 $I$-\magarc, as
determined from the data.  This
diagram shows clearly the selection biases against low surface
brightness and small scalesize galaxies.  Only the highest surface
brightness, largest scalesize galaxies can be seen out to 100\,Mpc.
The galaxies near the 100\,Mpc line have a 125 times larger visibility
volume than the galaxies near the 20\,Mpc line, which makes visibility
corrections essential to calculate real space density distributions
from the apparent distribution in the figure.

\section{SPACE DENSITY DISTRIBUTIONS}
\label{bivar}

Before we can calculate the true space density of galaxies using the
equations derived in \S\ref{viscor}, we have to determine the
uncertainty in the diameter selection parameter.  To this end, we
obtained 250 $B$-band images of galaxies in the ESO-LV catalog (Lauberts \&
Valentijn~\citeyear{LauVal89}), scanned from the same photographic
plates that were used to define the ESO-Uppsala catalog from which our
sample was selected.  One of us (RSdJ) went three times through the
images, measuring the diameters with a cursor on a computer screen.
These three sets of diameters were compared to the ESO-Uppsala
diameters and compared to each other.  It was found that the
uncertainty in the diameters was more constant in the absolute than
the relative sense in the range of diameters where we can be
reasonably sure that we are complete (2.2\arcmin$\le$\,$D_{\rm
maj}$\,$\le$4.2\arcmin).  The rms error between our measurements was
0.21\arcmin, while the rms error between our diameter measurements and
the ESO-Uppsala diameters was 0.31\arcmin.  This difference results
from the difference in measurement technique (with eye, magnifying
glass and ruler versus computer screen and cursor).  The ESO-Uppsala
diameters were quantified to the nearest 0.1 minute of arc, while the
human brain has a preference for `nice' numbers. The diameter
distribution of the ESO-Uppsala catalog shows distinct peaks at
2\arcmin, 2.2\arcmin, 3\arcmin, 3.5\arcmin, 4\arcmin\ and 5\arcmin.
If we had re-measured the ESO-Uppsala diameters in exactly the same
way as was done originally, we expect that the rms difference between our
own and the ESO-Uppsala measurements would have been lower than
determined now, and we therefore adopt an uncertainty in the
ESO-Uppsala diameters of 0.25\arcmin\ to be used in
equation\,(\ref{unccor}).

\ifthenelse{\value{emulate}=1}{
\begin{figure}[tb]
\epsfxsize=\linewidth \epsfbox[108 88 469 385]{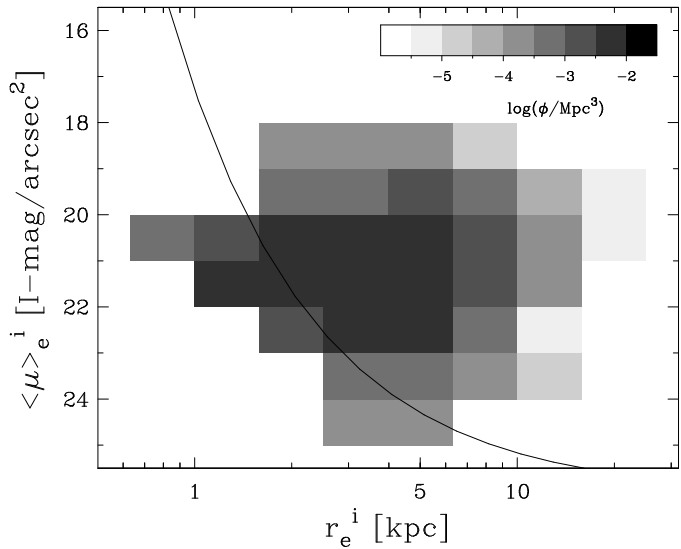}
\figcaption[f3.ps]{Bivariate space density distribution of Sb-Sdm galaxies as
a function of effective surface brightness and effective radius. The
line indicates the 20\,Mpc sample selection limit for 
face-on exponential disks. To the left of the line we are limited by
small number statistics and local density fluctuations.
\label{disre_muave}
}
\end{figure}
}{}

To calculate the true space density of galaxies in the (\muave,
\re)-plane, we have to weight each of the galaxies in
Fig.\,\ref{uncordis_re_muave} using the visibility correction
equations given in \S\ref{viscor}.  In Fig.\,\ref{disre_muave} we show the
space density of Sb-Sdm galaxies in number per Mpc$^3$. The 20\,Mpc
visibility limit of face-on galaxies with exponential disks is
indicated by the solid line.  To the left of this line we are limited
by small number statistics and local density fluctuations, but to the
right we should have a reasonably fair sampling of the local universe.
The limits on the distribution at the high surface brightness and
large scalesize ends are therefore real.  Note for instance that this
distribution strongly suggests that a galaxy like Malin~I (Bothun et
al.~\citeyear{Bot87}) with $\muavei \simeq 26$\,$I$-\magarc\ and 
$\re \simeq 140\,$kpc must be extremely rare.

\ifthenelse{\value{emulate}=1}{
\begin{figure}[tb]
\epsfxsize=\linewidth \epsfbox[108 88 469 385]{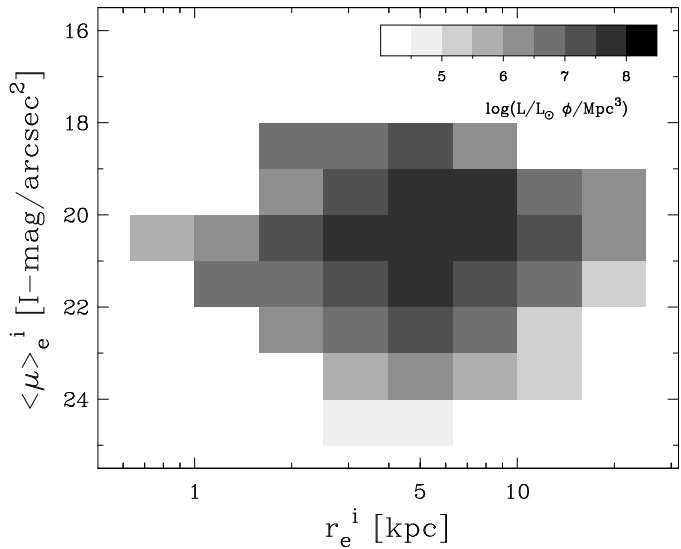}
\figcaption[f4.ps]{Bivariate luminosity density distribution of Sb-Sdm
galaxies as a function of effective surface brightness and effective
radius.
\label{disre_muave_magwei}
}
\end{figure}
}{}

Perhaps even more important than the space density of galaxies is
their luminosity density, which is presumably indicative of the
stellar (and baryonic) mass density.  Weighting each galaxy in
Fig.\,\ref{disre_muave} with its luminosity results in
Fig.\,\ref{disre_muave_magwei}, where we have used $M_{\sun} =
4.14$\,$I$-mag to convert the $I$-band
magnitudes to luminosities in solar units (Cox \citeyear{Cox00},
transfering from Johnson to Kron-Cousins $I$-band using Bessell
\citeyear{Bes79}).  Spiral galaxies with
effective radii of order 6\,kpc and $\muave\simeq 20$\,$I$-\magarc\
provide most of the spiral galaxy luminosity in the local universe.
It should come as no surprise that we live in a galaxy with these
qualifications.  The contribution of LSB spiral
galaxies to the total luminosity density of the universe apears to be
small. We will discuss this issue in more detail in \S\ref{discuss}.

\section{A FUNCTIONAL FORM}
\label{func}

In this section we will derive a functional form to describe the
bivariate distributions calculated in the previous section.  The
parametrization of the bivariate distributions will be useful to
compare distributions derived from differently selected samples and to
study redshift evolution.  The parametrization can also be used in
modeling where both galaxy luminosity and size are required (e.g.\
modeling the cross-sections of galaxies for producing quasar
absorption lines).

\ifthenelse{\value{emulate}=1}{
\begin{figure}[tb]
\epsfxsize=\linewidth \epsfbox[20 145 485 700]{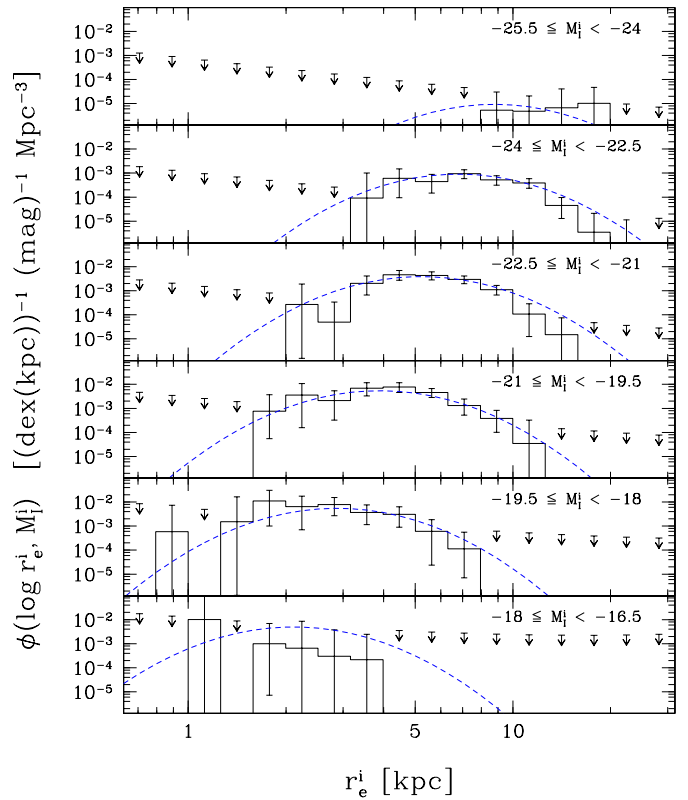}
\figcaption[f5.ps]{Bivariate space density distribution of Sb-Sdm galaxies as
a function of inclination corrected effective radius in different bins
of absolute $I$-band magnitude as indicated in the top-right corner of
each panel. The error bars on the histogram indicate the 95\%
confidence limits due to Poisson statistics and distance and diameter
uncertainties. The upper limits are the 95\% confidence upper limits
derived from non-detections assuming Poisson statistics for face-on
exponential disks of the given parameters. The dashed line shows the
fitted bivariate distribution function of equation\,(\ref{bivareq}) as
described in the text.
\label{bivar_magtot_re}
}
\end{figure}
}{}

In the previous sections we used the distributions in the
(\muave,\re)-plane, as these parameters are the most naturally
connected to the diameter selection limits.  In this section we will
instead use the distribution in the (\magtot,\re)-plane
(Fig.\,\ref{bivar_magtot_re}), as these quantities are the more
natural ones in the galaxy formation model we will to use to find
a suitable functional form for the bivariate distribution.  The two
descriptions are fully equivalent (except for some binning
differences) through the equation $\magtot = \muave - 5\log(\re) -
2.5\log(2\pi)$.

\subsection{Derivation of functional form}
\label{funcdev}

We will assume that the bivariate distribution can be written as the
product of the distribution in luminosity, assumed to be a Schechter
function, multiplied by a distribution in scalesize at a given
luminosity. To motivate a particular form for the latter, we consider
a simplified form of the Fall \& Efstathiou (\citeyear{FalEfs80})
disk galaxy formation model, as given by Fall (\citeyear{Fal83}). 

In the Fall \& Efstathiou (\citeyear{FalEfs80}) model, the scalesize of a
galaxy is determined by its angular momentum, which is acquired by
tidal torques from neighboring objects in the expanding universe,
prior to the collapse of the halo.
The total angular momentum of the system is usually expressed in terms
of the dimensionless spin parameter (Peebles \citeyear{Pee69})
\begin{equation}
 \lambda = J |E|^{1/2} M_{\rm tot}^{-5/2} G^{-1},
 \label{spin}
\end{equation}
with $J$ the total angular momentum, $E$ the total energy and $M_{\rm
tot}$ the total mass of the system, all of which are dominated by the
DM halo.  N-body simulations (e.g.\ Barnes \&
Efstathiou~\citeyear{BE87}; Warren et al.~\citeyear{War92}) show that
the distribution of $\lambda$ values of DM halos acquired from tidal
torques in hierarchical clustering cosmologies can be well be
approximated by a log-normal distribution
 \begin{equation}
 P(\lambda)\,d\lambda = \frac{1}{\sqrt{2\pi}\sigma_{\lambda}}
	\exp\left[ -\frac{\ln^2(\lambda/\lambda_{\rm med})}{2\sigma^2_{\lambda}}\right]
	\,\frac{d\lambda}{\lambda}
 \label{spineq}
 \end{equation}
The median $\lambda_{\rm med}$ and dispersion (in $\ln\lambda$)
$\sigma_{\lambda}$ are found to depend remarkably weakly on the
cosmology, halo mass or initial spectrum of density fluctuations
(e.g.\ Barnes \& Efstathiou~\citeyear{BE87}; Warren et
al.~\citeyear{War92}; Cole \& Lacey~\citeyear{ColLac96}), with typical
values $\lambda_{\rm med}\approx 0.04$ and $\sigma_{\lambda}\approx
0.5-0.6$.

With some simplifying assumptions, we can now relate the halo parameters
in the definition~(\ref{spin}) to the disk radius and luminosity.  (i)
We model the halo as a singular isothermal sphere (density $\propto
1/r^2$), with circular velocity $V_c$ and total mass $M_{\rm tot}$. 
From the virial theorem we then obtain $E \propto V_c^2 M_{\rm
tot}$.  (ii) We assume that the galaxy is a perfect exponential disk,
with (baryonic) mass $M_D$ and effective radius \re.  We also assume
that the disk circular velocity is equal to that of the halo (i.e.\ we
ignore the self-gravity of the disk).  The disk angular momentum then
scales as $J_D  \propto M_D \re V_c$.
(iii)
We assume that the specific angular momentum of the disk is proportional
(or equal) to that of the halo $J_D/M_D \propto J/M_{\rm tot}$.  (iv) We also
assume that the ratio of baryonic to dark matter is constant, and that
the same fraction of the baryonic mass always ends up in the disk,
resulting in disk mass being proportional to halo mass $M_D \propto
M_{\rm tot}$.  
Combining these results in equation~(\ref{spin}), we find $\lambda
\propto \re V_c^2/M_D$.  We now want to express this in terms of the
disk luminosity $L$.  (v) We assume a power law relation between disk
mass and luminosity: $M_D\propto L^\gamma$, with $\gamma$ expected to
be close to 1.  The power $\gamma$ incorporates the effect of
variations in $M_D/L$ due stellar population differences (de Jong
\citeyear{deJ4}; Bell \& de Jong \citeyear{BeldeJ00a}) and to
variations in gas mass fractions (McGaugh \& de Blok
\citeyear{McGdeB97}), which tend to be functions of surface brightness
and $L$.  (vi) Finally, we use the observed Tully \& Fisher
(\citeyear{TulFis77}) relation $L\propto V_c^\epsilon$, with
$\epsilon\sim 3$ in the $I$-passband.  These approximations yield
$\lambda \propto \re L^{[(2/\epsilon)-\gamma]} \simeq \re L_I^{-1/3}$.
As an alternative to step (vi), we could use the relation $M_{\rm tot}
\propto V_c^3$ predicted for DM halos, assuming that they all have the
same mean density.  This leads to $\lambda \propto \re L^{-\gamma/3}
\simeq \re L_I^{-1/3}$, in practice very similar, but relying more on
theory than observations.  Both cases can be
written as $\lambda \propto \re L_I^{\beta}$, with $\beta\simeq -1/3$.

As $\lambda$ is expected to have a log-normal behavior, this means
that, {\em at a given luminosity, this simple form of the Fall \&
Efstathiou model predicts the distribution of scale
sizes to be log-normal, and the median value of \re\ to vary with
luminosity as $\re \propto L^{-\beta} \sim L^{1/3}$}.  Combining this
result with the Schechter LF, the full bivariate function for space
density as function of luminosity and effective radius becomes:
\begin{eqnarray}
 \label{bivareql}
\frac{d^2n}{dL\ d\re}\, &dL&\ d\re = 
\phi_* \left( \frac{L}{L_*} \right)^{\alpha}
\exp\left(-\frac{L}{L_*} \right)\, \frac{dL}{L_*} \\ 
& \times &  \frac{1}{\sqrt{2\pi}\sigma_{\lambda}} 
\exp\left[ -\frac{\ln^2\left( (\re/\res)(L/L_*)^{\beta} \right)} 
{2\sigma^2_{\lambda}} \right]\, \frac{d\re}{\re} \nonumber
\end{eqnarray}
This can be rewritten in terms of absolute magnitudes ($M$) as
\begin{eqnarray}
 \label{bivareq}
%
%
 \lefteqn{\phi(M,\log(\re))\,dM\,d\log\re =}  \\
 &   & 0.4\ln(10)\ \phi_*\ 10^{-0.4(\alpha+1)(M-M_*)} \,
     \exp(-10^{-0.4(M-M_*)}) \,dM \nonumber \\[1.3mm]
 & \times & \frac{\ln(10)}{\sqrt{2\pi} \sigma_\lambda} 
     \exp\left[ -\frac{1}{2}\left( \frac{\log(\re/\res)-0.4\beta(M-M_*)}
     {\sigma_\lambda/\ln(10)} \right)^2 \right] \, d\log\re \nonumber
 \end{eqnarray}
where absolute magnitude $M_*$ corresponds to luminosity $L_*$.

The first line in equations~(\ref{bivareql}) and (\ref{bivareq}) is
the Schechter LF and the second line represents the log-normal scale
size distribution at a given luminosity. In these equations,
$\phi_*$, $\alpha$ and $M_*$ (or $L_*$) have the usual meanings for a
Schechter LF, while $\res$ gives the median disk size for a galaxy
with $M=M_*$, and $\beta$ the slope of the dependence of the median
\re\ on $L$.  The quantity $\sigma_\lambda$, which was defined in
equation\,(\ref{spineq}) as the dispersion in $\ln(\lambda)$, is shown
in equations~(\ref{bivareql}) or (\ref{bivareq}) to equal the
dispersion in $\ln(\re)$ at a given luminosity.  Note that this
function is slightly different from de Jong \& Lacey
(\citeyear{deJLac00}), as we have taken the factor $\ln(10)$ out of the
scalesize normalization. This function is identical in shape to the
one used by Cho\l oniewski (\citeyear{Cho85}) to describe the
bivariate distribution function of E and S0 galaxies.

\ifthenelse{\value{emulate}=1}{
\begin{figure*}[t]
\epsfxsize=\linewidth \epsfbox[27 165 582 495]{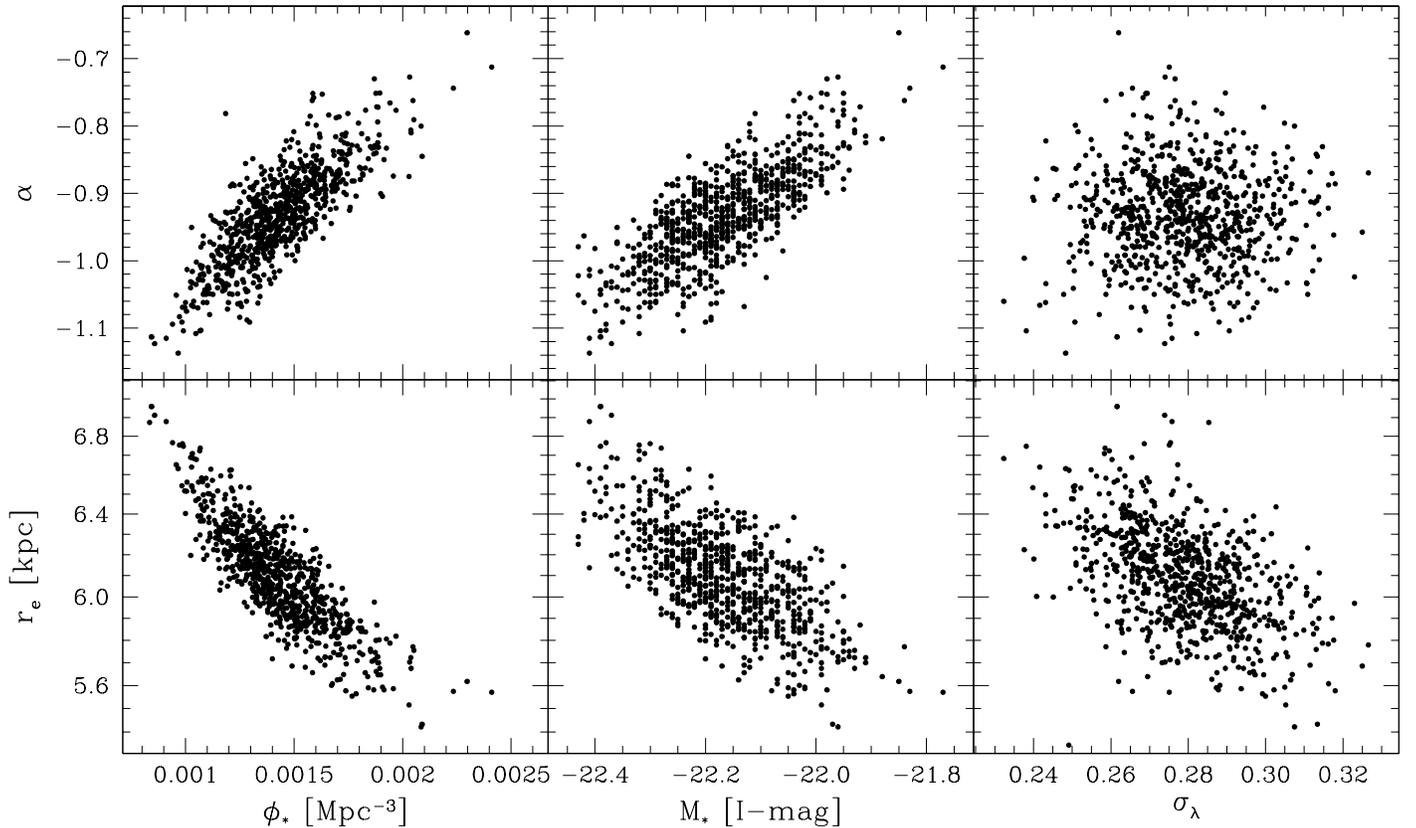}
\figcaption[f6.ps]{Correlations in the bootstrap resampled fitted parameters of
equation\,(\ref{bivareq}), for the case of total galaxy magnitudes.
\label{disparbivar}
}
\end{figure*}
}{}

The simple model that we used to derive equation~(\ref{bivareql}) (or
\ref{bivareq}) ignored some important aspects of the physics of galaxy
formation, and furthermore the Schechter LF was simply assumed based
on observations, rather than being derived from theory (although the
form of the Schechter LF for galaxies was originally inspired by the
mass function of DM halos in hierarchical clustering models derived by
Press \& Schechter (\citeyear{PreSch74})). Each of the assumptions
(i--vi) used to derive equations ~(\ref{bivareql}) \& (\ref{bivareq})
carry their own uncertainties. Most notably, if galaxies are built up
mainly by merging of baryonic sublumps rather than by smooth accretion
of gas, as found in many numerical simulations (e.g.\ Navarro, Frenk \&
White \citeyear{Nav95}), then the baryons may lose most of their
initial angular momentum. There may not be a one-to-one
correspondence between disk and halo angular momenta, violating
assumption (iii), although, a correlation between the angular momenta
is still expected, albeit with much scatter (e.g.\ Navarro \& Steinmetz
\citeyear{NavSte00}). However, in this case, galaxy disks are also
predicted to have much smaller radii than is observed. Suppression of
early cooling of the gas by feedback from supernovae may be able to
prevent this process of drastic angular momentum loss (e.g.\ Weil, Eke
\& Efstathiou \citeyear{Weil98}, Sommer-Larsen, Gelato \& Vedel
\citeyear{Somm99}), and rescue our general model for disk
formation. The assumption (iv) of a constant ratio of disk to halo
mass when combined with the assumption in (vi) that $M_{\rm tot}
\propto V_c^3$ predicts a ``baryonic'' Tully-Fisher relation $M_{\rm
D} \propto V_c^3$. This may conflict with observations: McGaugh et
al.~(\citeyear{McG00}) find a slope close to 4 rather than 3 (but see
Bell \& de Jong \citeyear{BeldeJ00b} who argue it is less than 3.5).  
This problem can be (partly)
resolved if the observed rotation velocity $V_c$ is not the same as
the DM halo rotation velocity (van den Bosch \citeyear{vdB00}, but see
Mo \& Mao \citeyear{MoMao00}) or if the baryon-to-DM fraction changes
systematically with $V_c$.  We return to the simplifications
and uncertainties in these kind of models in \S\ref{sams}, where we
consider the predictions from semi-analytic models of galaxy
formation, which include much more detailed physical treatments of the
evolution of both DM and baryons than we did here, and relax some of
the assumptions. For the moment, our
derivation suffices to motivate the use of equation~(\ref{bivareq}) in
fitting observational data.


\subsection{Fitting the data}

Before we can fit equation\,(\ref{bivareq}) to the data, we have to
understand the uncertainties in the data points. As mentioned in
\S\ref{viscor}, the errors on the \vmax-corrected data points tend
to be dominated by Poisson statistics. Especially in bins where we
have few galaxies, these errors are highly asymmetric, and we cannot
use a simple $\chi^2$ minimization method to fit the data. The 95\%
confidence limits that we plot on the histograms of
Fig.\,\ref{bivar_magtot_re} were calculated taking into account
the distance and diameter uncertainties as described in \S\ref{viscor} and the
Poisson confidence limits (as described by e.g.\ Gehrels
\citeyear{Geh86}).

In addition, the bins with no galaxy detections also carry information
which we can use to fit our parameterization to the data.  We can
calculate for a galaxy with given structural parameters the detection 
volume \vmax\ and set an upper limit to the number of galaxies with
these structural parameters in that volume.  To calculate the upper
limit to the number of galaxies in a bin in the (\magtot,\re)-plane, we
have to calculate the \vmax\ of a galaxy with parameters (\magtot,\re). 
We therefore have to link (\magtot,\re) to our selection limits $D_{\rm
max}$ and $D_{\rm min}$.  We determined the average surface brightness
of our galaxies at their major axis diameters $D_{\rm maj}$.  For a
face-on exponential disk with given \magtot\ and \re\ we can now
calculate the diameter at this surface brightness, and hence the minimum
and maximum visible distances, and so \vmax.  The surface brightness
at $D_{\rm maj}$ of the galaxies showed a rather large spread, and a
small dependence on the effective surface brightness of the galaxies,
which was taken into account when calculating the non-detection \vmax. 
A non-detection in Poisson statistics gives a 95\% confidence upper
limit of 2.996 galaxies on the true average number of galaxies in the
corresponding \vmax\ (Gehrels \citeyear{Geh86}), which are the upper limits
plotted in Fig.\,\ref{bivar_magtot_re}. 

\begin{table*}[tbh]
 \caption{Bivariate distribution function parameters.} 
{
\tabcolsep=0.39cm
\begin{tabular*}{\textwidth}{lcccccc}
\tableline
\tableline
\multicolumn{1}{c}{fit}   & $\phi_*$ & $\alpha$ & $M_*$ & \res & $\sigma_\lambda$ & $\beta$\\
       &Mpc$^{-3}$& & $I$-mag & kpc & & \\
\tableline
total galaxy & 0.0014$\pm$0.0003 & -0.93$\pm$0.10 & -22.17$\pm$0.17 & 6.09 $\pm$ 0.35 & 0.28$\pm$0.02 & -0.253$\pm$0.020\\
disk only    & 0.0014$\pm$0.0003 & -0.90$\pm$0.10 & -22.38$\pm$0.16 & 5.93 $\pm$ 0.28 & 0.36$\pm$0.03 & -0.214$\pm$0.025\\
\tableline
\end{tabular*}
}
\tablenotetext{}{\ \vspace*{-0.9cm}\\%
All errors are 95\% confidence limits as determined
from bootstrap resampling.}
 \label{fittab}
\end{table*}

We used maximum likelihood fitting to determine the parameters in the
bivariate distribution function.  Initial estimates for the parameters
were obtained with a non-linear $\chi^2$ minimization routine based
on the Levenberg-Marquardt method, which were used as a starting
point for the downhill simplex method (Press et al.~\citeyear{Pre93}) used
to implement the maximum likelihood fitting. We used only the Poisson
distribution to calculate the likelihood distribution in each bin,
which was minimized in the negative log (see also Cash \citeyear{Cas76}):
\begin{equation}
\log(P) = \sum_i^{N_{\rm bin}} x - N^i * \log(x)
 \end{equation}
 \begin{equation}
x = \max(1,N^i)\,\phi^i_{\rm mod}/\phi^i_{\rm obs}
 \end{equation}
 where we sum $i$ over all ${N_{\rm bin}}$ bins (also the bins with no
galaxies), having $N^i$ galaxies per bin. $\phi^i_{\rm mod}$ is
the predicted space density of objects in our model from
equation\,(\ref{bivareq}), and $\phi^i_{\rm obs}$ is the observed space
density  in bin $i$ calculated as described in \S\ref{bivar}, or, if
the bin contains no observed galaxies, the value of $1/\vmax$
calculated for the 
upper limit.  In general the upper limits hardly influence the fit,
unless the fit function approaches very close to the upper limits.  We did
not use the distance uncertainties in the calculation of the probability
distribution, as the errors are dominated in all bins by Poisson small number
statistics.  To match the data, we binned the model function
by integration over the same bin ranges as the data. 

We used bootstrap resampling to estimate the errors on the bivariate
distribution function parameters (Press et al.~\citeyear{Pre93}).  For
each bootstrap sample, the same total number of galaxies were randomly
selected from the original sample (meaning some galaxies were selected
several times, others not at all), and the whole analysis and
parameter fitting was performed again.  This bootstrap resampling was
performed 50 times.  Even though we binned the model function in the
same way as the data, the fit parameters depended slightly on bin
sizes.  Therefore the whole bootstrap analysis was performed on 4
different steps in \magtot\ bin size and 4 steps in \re\ bin size,
resulting in 800 independent parameter measurements.  The
distributions of these points for some of the parameters are plotted
in Fig.\,\ref{disparbivar}.

Table\,\ref{fittab} lists the fit results for two cases, one for
\magtot\ and \re\ determined for the full galaxy (including the bulge)
and one for the disk only.  The errors in the Schechter LF parameters
are strongly correlated as usual (see Fig.\,\ref{disparbivar}), so the
95\% confidence limits indicated in Table\,\ref{fittab} are strongly
correlated for $\phi_*$, $\alpha$, $M_*$ and $\res$.  The width of the
scalesize distribution at a given magnitude, as parameterized by
$\sigma_\lambda$, is rather uncorrelated with the other parameters and
is well defined.  The value of $\sigma_\lambda\simeq0.28$ for the
total galaxies and $\sigma_\lambda\simeq0.36$ for the disks only is
rather smaller than the $\sigma_\lambda=0.5-0.6$ typically found from
cosmological N-body simulations.  Some possible explanations will be
discussed in \S\,\ref{discuss}.

We find that our Malmquist edge bias correction does significantly
change our results. Not correcting for edge bias increases $\alpha$ by
about 0.1, with the other parameters changing according to the trends
of the bootstrap resampled scatter diagrams of
Fig.\,\ref{disparbivar}. Therefore, to obtain an accurate
determination of the LF, it is important to have small errors in the
galaxy selection parameters.

The values we find for the Schechter LF parameters are very similar to
other recent LF determinations of spiral galaxies. Marzke et
al.~(\citeyear{Mar98}) find for example $\phi_*=0.0022$, $\alpha=-1.1$
and $M_*=-22.07$ (converting to our $H_0$ and using $B$--$I$=1.7\,mag
(de Jong~\citeyear{deJ4})). Marinoni et al.~(\citeyear{Mar99}) find for
spiral galaxies $\phi_*\sim 0.0016$, $\alpha\sim -0.85$ and $M_*\sim
-22.4$ (averaging the Sa-Sb and the Sc-Sd determinations). The fact
that these values are so similar suggest that there is no huge
population of Sb-Sdm LSB galaxies, as our determination does correct
for the bias against LSB galaxies, while this is not the case for the
other studies. We will address this point in more detail in
the next section.

\section{DISCUSSION}
\label{discuss}

We have shown that the space density distribution of spiral galaxies
can be described by a Schechter LF in luminosity combined with a
log-normal scalesize distribution at a given luminosity.  We use the
goodness-of-fit parameter $Q$ (Press et al.~\citeyear{Pre93}) to
determine how well our function is fitting the data.  $Q$ indicates
the probability that the measured $\chi^2$ is exceeding the expected
$\chi^2$ by chance, given the number of degrees of freedom.  Normally
a $Q>0.1$ is accepted as a good fit and $Q>0.001$ is acceptable when
the errors are not normally distributed.  We find $Q>0.1$ in 57\% of
the bootstrap resampled realizations of the data, and $Q>0.001$ in
more than 95\% of the realizations.  This is a remarkably good result
considering that (1) we have not fit to the minimum in $\chi^2$,
instead using our maximum-likelihood technique to take into account
the non-Gaussian error distribution on the data points, and (2)
outliers are more likely to occur because our errors are not normally
distributed.  Indeed, the smallest Q values occur when we have a fine
binning in magnitude and/or scalesize, so that the number of bins with
few galaxies increases and the errors become very asymmetric and
non-Gaussian.  

We conclude that our parameterization gives an accurate description of
the observed bivariate distribution given the known
uncertainties. This conclusion holds true, independent\-ly wheth\-er one
believes in its derivation based on a particular model for disk
formation,  and despite
the known simplifications and uncertainties in the derivation. This
does of course not mean that this function is unique in giving a good
description. Especially with better number statistics a more detailed
model may be necessary. Some hint of this can already be seen in
Figs.\,\ref{bivar_magtot_re} \& \ref{moddis}, where the scalesizes of
the galaxies in the brightest magnitude bin seem to be larger than
modeled by the function.

\ifthenelse{\value{emulate}=1}{
\begin{figure}[tb]
\epsfxsize=\linewidth \epsfbox[33 172 527 537]{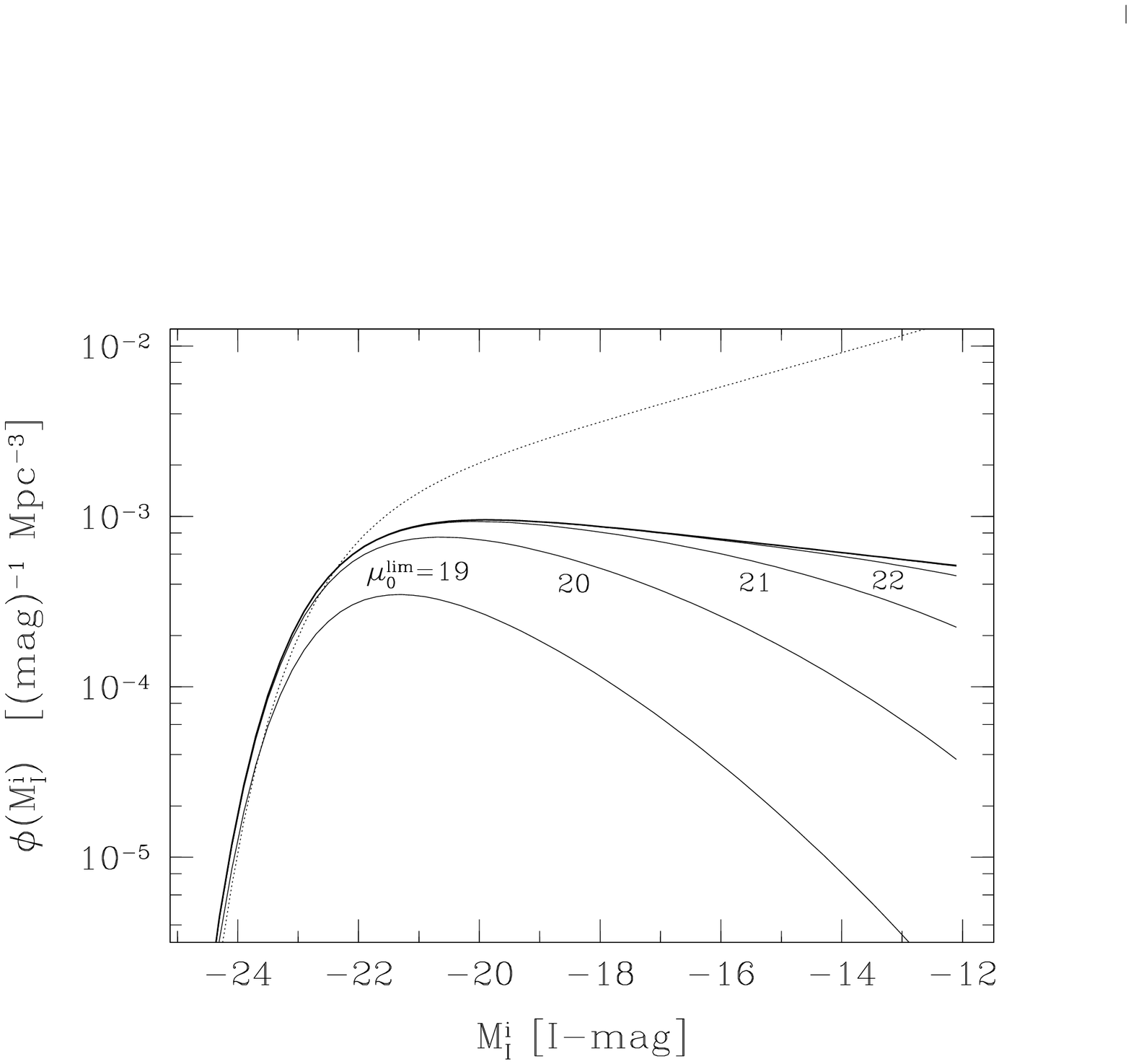}
\figcaption[f7.ps]{Galaxy disk luminosity functions derived from our disk
only bivariate distribution parameterization.
The thin solid lines are LFs limited to galaxies with
disk central surface brightnesses brighter than the indicated values
in $I$-\magarc. The thick line at the top is the disk LF integrated
over all surface brightnesses. To indicate how inclusion of galaxies
of type later than Sdm might influence this diagram (and
Figs.\,\ref{mudis_maglim} and \ref{phifuncmuMwei}) we also show a LF with
$\alpha=-1.25$, leaving all other parameters in our bivariate
distribution the same (dotted line; for details see text).
\label{LFdis_cslim}
}
\end{figure}
}{}

\subsection{One Dimensional Projections}
\label{1Dproj}

Given our 2D parameterization, we will now investigate some 1D
projections of this parameterization, and determine how these 1D
projections depend on limits placed on one of the other parameters.
Unfortunately, we cannot use the real data to make these
projections. Due to selections limits, there are regions in the 2D
plane where we have no data, only upper limits. A 1D integration would
mainly look like a meaningless upper and lower limits plot. By using
the 2D parameterization,  we assume that the same function that
fits in the observed region also describes galaxies in the regions where
we have no data.  In this section we use the disks only parameterization of
\S\ref{func}.

In Fig.\,\ref{LFdis_cslim} we show how limits on the surface
brightness in a sample can influence the determination of the LF.  We
integrate the bivariate distribution function down to the central
surface brightness limits indicated in Fig.\,\ref{LFdis_cslim}, thus
calculating the LF for all galaxies with central surface brightness
brighter than the indicated limits.

For local samples selected from photographic plates, the number of low
surface brightness Sb-Sdm galaxies missing is expected to be quite
small.  For instance, the ESO-Uppsala Catalog of Galaxies (Lauberts
\citeyear{Lau82}) has a typical surface brightness at the selection
diameter of about 24.8 $I$-\magarc\ as determined from this sample, while the
Uppsala Galaxy Catalog (UGC; Nilson \citeyear{UGC}) has a selection
surface brightness of about 26\,$B$-\magarc\ (de Jong \& van der Kruit
\citeyear{deJ1}), which corresponds to about 24.3 $I$-\magarc\ with 
$B$--$I\simeq$1.7 (de Jong \citeyear{deJ4}). So requiring that the
{\em central} surface brightness of the galaxies are at least 1 mag
brighter than the selection surface brightness in order to be selected
(very generous considering that bulges make detection even easier),
the ESO-Uppsala catalog is expected to be reasonably complete down to
a central surface brightness of \muo$\simeq$23.8 $I$-\magarc, the UGC down to
\muo$\simeq$23.3 $I$-\magarc. Figure\,\ref{LFdis_cslim} therefore indicates
that LFs determined from local samples selected from POSS-like
photographic plates should be reasonably complete to
$M_I\simeq-14$\,mag.

The situation regarding the surface brightness bias for galaxy samples
of types later than Sdm is less clear. It is well established that the
latest type (i.e.\ irregular and dwarf) galaxies have a much steeper
faint end slope of the LF than the spiral galaxies studied here
(Marzke et al.~\citeyear{Mar98}; Lin et al.~\citeyear{Lin99}). Our
Schechter LF slope agrees well with slopes of other pure spiral galaxy
samples. For galaxy samples of later types the slope is much steeper,
and as surface brightness and luminosity are correlated in our
parametrization, the number of missing LSB galaxies will increase when
one considers late types.

We can get some feeling for how many galaxies of types later than Sdm
we are missing by comparing the number of galaxies of types 7--8 in
the ESO-Uppsala catalog to the number of type 9--10 galaxies. For the
diameter and inclination selection criteria we applied to define our
sample we have about twice as many $T=7-8$ galaxies as $T=9-10$
galaxies. When we use the full ESO-Uppsala catalog, the numbers are
about equal. This suggests that type 9--10 galaxies are typically
smaller and of lower surface brightness than the type 7--8 galaxies,
which are already the smallest type of galaxies included in our sample
(see Fig.\,\ref{uncordis_re_muave}). We do not have the photometry to
make the full bivariate correction, but for the galaxies with
redhsifts, we can make a \vmax\ comparison to estimate relative
number densities. Using NED we obtained redshifts for 90\% of the
$T=7-8$ and 80\% of the $T=9-10$ galaxies with 
$D_{\rm maj}>1.65\arcmin$ (63\% and 44\% respectively for the full
sample). For the sample with a $D_{\rm maj}>1.65\arcmin$ cut-off, we
then find that the \vmax\--corrected number density of $T= 9-10$
galaxies is about 17 times as high as that of the $T=7-8$ galaxies in
our sample. For the full ESO-Uppsala catalog, the volume density of
type 9--10 galaxies is about 25 times that of the type 7--8 galaxies.

These relative space densities are rather uncertain due to redshift
incompleteness and due to the generally low redshifts of these
galaxies, making Hubble flow distances rather uncertain. It does
however show that a considerable number of disk galaxies are not covered by
this study, in particular at the faint end of the LF. It is not
inconceivable that this effect will raise the slope of the faint end
of the combined LF of types 3--8 and 9--10 galaxies by a few
tenths. In the remainder of this section we investigate the effect of
including type 9--10 galaxies by also showing 1D projections with
faint end slopes with $\alpha=-1.25$, leaving all other parameters in
the bivariate function the same. 

The $\alpha=-1.25$ parameterization is rather ad hoc and is only
intended to give an indication of what including dwarf galaxies might
do to the 1D projections. We have no way of knowing whether these
late-type galaxies follow the same distribution of scalesize with luminosity
as earlier-type spiral galaxies do, nor about the exact value
for the faint end slope of the LF. The $\alpha=-1.25$ value for the
faint end slope is inspired by some recent determinations of the LF
where late-type galaxies have explicitly been included (e.g.\ Marzke et
al.~\citeyear{Mar98}, Zucca et al.~\citeyear{Zuc97}, Folkes et
al.~\citeyear{Fol99}). The $\alpha=-1.25$ parameterization is almost
certainly too simple according to these studies, as the very late type
galaxies have an LF with a very steep faint end slope, but also only
become significant in number density at very faint luminosities. Therefore
in reality the LF may steepen at very low luminosities, rather than
being described by a single Schechter function, but it
is beyond the scope of this paper to investigate the effects of this. 

The ``dwarf corrected'' $\alpha=-1.25$ LF is shown as a dotted line in
Fig.\,\ref{LFdis_cslim}. For bivariate distributions with steeper
faint end slopes the incompleteness due to surface brightness limits
quickly becomes more severe.  For example, for a bivariate distribution
function with $\alpha=-1.25$ we start to underestimate the LF by a
factor of 2 at \magtot$=-14$\,mag if our surface brightness cut-off is at
22\,$I$-\magarc.

The selection against LSB galaxies can quickly become significant at high
redshifts due to the \zq\ redshift dimming. Not using a full bivariate 
distribution description in the comparison with local samples can give 
the false impression of evolution in the structural parameters of
galaxies. Surveys for high redshift galaxies will normally have much
lower surface brightness selection limits, but even for the
Hubble Deep Fields (HDFs), with their very low surface brightness limits of
29\,$I$-\magarc, the effects of surface brightness thresholds are
predicted to be significant at high redshifts, {\em if the galaxy
population does not evolve}.  Consider, for example, galaxies detected
as $U$-band dropouts at a redshift of about 3.  These galaxies suffer
6 magnitudes of surface brightness dimming and about 2 magnitudes of
dimming due to the K-correction for an (unevolving) Sb galaxy.  If we
require that the central surface brightness of the galaxy has to be 1
magnitude above the sky to enable detection, the $z$=3 galaxy has to
have a rest-frame disk central surface brightness of about
19\,$I$-\magarc\ to be detected in the HDFs.  Such a surface
brightness cut-off would start to severely affect our determination of
the LF, even at $M_*$, as can be seen in Fig.\,\ref{LFdis_cslim}.
Luckily, this approach is probably overly pessimistic, as central
bulges may raise the central surface brightness to help detection,
and galaxy evolution will make the galaxies bluer at high redshift and
hence reduce the K-correction.  In addition, hierarchical galaxy
formation models predict that the galaxies existing at high redshift
should typically have smaller radii than present-day galaxies, which
will also tend to increase their surface brightnesses. Still,
comparisons of structural parameters at different redshifts will
require determinations of bivariate distribution functions to take
into account varying selection functions.

\ifthenelse{\value{emulate}=1}{
\begin{figure}[tb]
\epsfxsize=\linewidth \epsfbox[33 172 527 537]{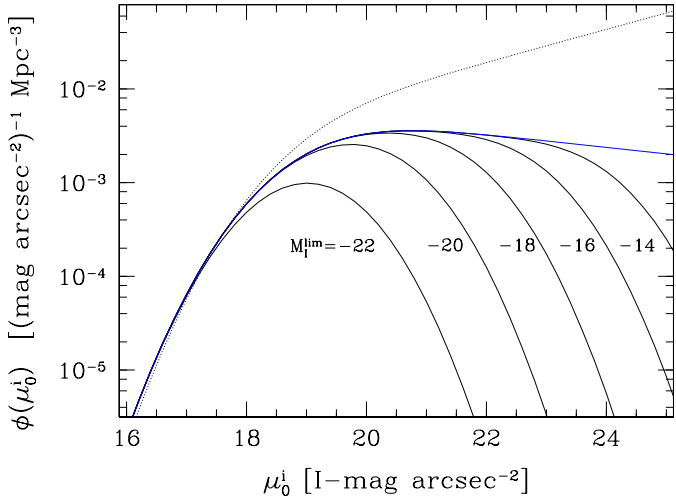}
\figcaption[f8.ps]{The distribution of the disk central surface brightnesses
for all galaxies brighter than the indicated absolute $I$-mag, derived 
from the disk only bivariate distribution function parametrization. The
thick line at the top indicates the surface brightness distribution
for all galaxies, assuming that the LF continues with the same slope
to the faintest magnitudes. The dotted line shows the integrated disk
central surface brightness distribution for all galaxies when we
change $\alpha$ in the bivariate distribution function from -0.90 to
-1.25, leaving all other parameters the same, to simulate what
inclusion of galaxies of type later than Sdm might do to this
distribution.
\label{mudis_maglim}
}
\end{figure}
}{}

In Fig.\,\ref{mudis_maglim} we show the distribution of central
surface brightnesses integrated down to the indicated limiting
absolute magnitudes.  The slope at the faint end of the distribution
is determined by the faint end slope $\alpha$ of the Schechter LF and
the rate of change of median central surface brightness as function of 
luminosity as parametrized by $\beta$. The slope in magnitudes becomes
$d\log(\phi(\muo))/d\muo=-0.4(\alpha+1)/(2\beta+1)$, which is about
$-0.07$. Late-type galaxies
are expected to have a steeper faint end slope of their surface
brightness distribution, because they have a steeper LF.  We show an
estimate of the possible
size of this effect by the dotted line in Fig.\,\ref{mudis_maglim},
which shows the same bivariate function as before, except that
$\alpha$ has been changed from -0.90 to -1.25, i.e.\ assuming that the
late-type dwarfs missing from our observed sample follow the same
distribution of radius or surface brightness as a function of
luminosity as the spiral galaxies for which we have measured the
bivariate distribution function.

Our surface brightness distribution for spirals is somewhat similar to
the distribution presented by McGaugh et al.~(\citeyear{McG95}), even
though obtained by a completely different method.  In order to derive
their distribution from observations, they had to assume that surface
brightness is independent of scalesize (or more accurately, that the
shape of the scalesize distribution does not depend on surface
brightness), which is reasonably correct for the range of surface
brightnesses we have investigated (see e.g.\
Fig.\,\ref{disre_muave}). Their surface brightness distribution cuts
off at the bright end more steeply and at a fainter magnitude than
ours, which could be partly due to the different correction for selection effects
or to the use of $B$-band photometry uncorrected for
dust extinction.  Also O'Neil and Bothun (\citeyear{ONeBot00}) find a slowly
declining surface brightness distribution, doing a correct (though
relative, not absolute) \vmax\ correction. Unfortunately, the authors
of both investigations fail to indicate the exact range in scalesize
and/or magnitude their surface brightness distributions apply to and
direct comparisons are therefore impossible.

The number of LSB galaxies that our bivariate distribution function
predicts is somewhat lower than what has been found in surveys for LSB
galaxies. Dalcanton et al.~(\citeyear{Dal97a}) find a number density
of 0.022$\pm$0.011\,Mpc$^{-3}$ for galaxies with
23$<\muo<$25\,$V$-\magarc\ and $\red>0.78$\,kpc, while our bivariate function
predicts about 0.0065\,Mpc$^{-3}$ (using $V$--$I\sim1$\,mag (de
Jong~\citeyear{deJ4}) and correcting for the different $H_0$).
Sprayberry et al.~(\citeyear{Spr97}) find a number density of about
0.07\,Mpc$^{-3}$ for galaxies with 22$<\muo<$25\,$B$-\magarc, where
our function gives about 0.012 galaxies per Mpc$^{3}$ for this surface
brightness range. These discrepancies of a factor 4-6 in number
density seem rather large, but if we were to use a bivariate function
with $\alpha=-1.25$ to correct for the missing dwarf and irregular
galaxies then our bivariate function would be fully consistent with these
LSB surveys.

Tully \& Verheijen (\citeyear{TulVer97}) have argued that the central
surface brightness distribution of spiral galaxies is bimodal, in
particular when using $K$-band data.  We do not see such bimodality,
independent of whether we use their proposed bimodal dust extinction
correction, whether we use only the 200 most face-on galaxies with the
smallest extinction corrections, or whether we use bulge/disk
decomposed parameters or effective total galaxy parameters.  In the
many ways in which we have looked at the MFB data set, where we have
tried to minimize the effects of extinction and hence the difference
between $I$ and $K$-band, we have never
seen any bimodality in the surface brightness distributions.  Whether
the bimodal effect is the result of the special Ursa Major cluster
environment that was studied by Tully \& Verheijen (even though a fair
fraction of the MFB galaxies must lie in the outer parts of clusters)
or an unlucky case of small number statistics (Bell \& de Blok
\citeyear{BeldeB00}) remains to be seen.

\ifthenelse{\value{emulate}=1}{
\begin{figure}[tb]
\epsfxsize=\linewidth \epsfbox[33 172 588 537]{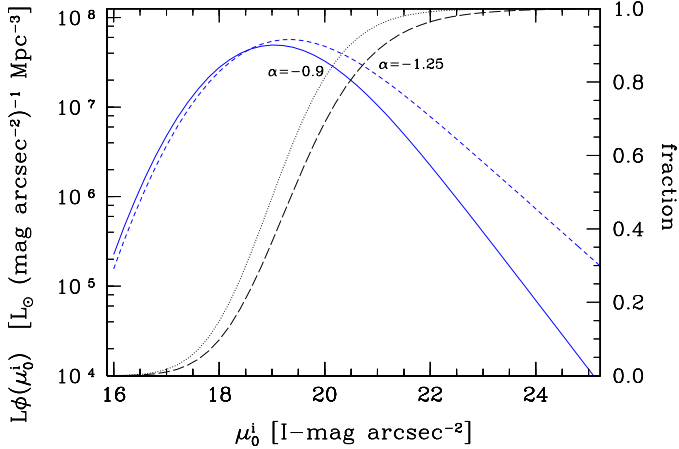}
\figcaption[f9.ps]{The differential (solid and short dashed lines; left axis) and
cumulative (dotted and long dashed lines; right axis) luminosity density of
disks as functions of disk central surface brightnesses. The solid and
dotted lines indicate the distributions for the standard bivariate
distribution function for disks
(Table\,\ref{fittab}). The short and long dashed lines are for the
same distribution, but with the faint end slope of the LF changed from
$\alpha=-0.90$ to $-1.25$ to simulate inclusion of galaxies of type
later than Sdm.
\label{phifuncmuMwei}
}
\end{figure}
}{}

The final 1D projection we are interested in is the luminosity density
of the local universe as a function of disk central surface brightness
as shown in Fig.\,\ref{phifuncmuMwei}. The thick solid line indicates
the luminosity density for our disk bivariate distribution function,
assuming the faint end of the LF continues for ever with the same
slope. Most of the luminosity density of Sb-Sdm galaxies is provided
by galaxies of $\muo\sim 19-19.5$\,$I$-\magarc.  Changing $\alpha$
from -0.90 to -1.25 to incorporate the effect of dwarfs and irregulars
changes that to slightly fainter surface brightnesses (thick short
dashed line). Looking at the cumulative distribution for Sb-Sdm
galaxies (thin dotted line), we see that 90\% of the spiral galaxy
luminosity in the local universe is provided by galaxies with
\muo\,$<\,20.5\,I$-\magarc. The 90\% level changes to
\muo\,$<\,21.2\,I$-\magarc\ when we use the $\alpha=-1.25$
parametrization. This corresponds roughly to 22.2 and
22.9\,$B$-\magarc\ respectively, using $B$--$I\sim$1.7\,mag for late
type galaxies (de Jong \citeyear{deJ4}). The faint end slope of the
combined spiral and dwarf/irregular LF would need to be significantly
steeper than -1.25 for LSB galaxies to become significant contributors
to the luminosity density of the local universe.
 
An earlier determination of the contribution of LSB galaxies to the
luminosity density of the local universe was presented by by
McGaugh~(\citeyear{McG96}). He estimated that 10-30\% of the local
luminosity density came from galaxies with central surface brightnesses
fainter than 22.75\,$B$-\magarc\ (i.e.\ $\sim 21\,I$-\magarc). We find
for the same cut-off about 4\% (about 12\% if $\alpha=-1.25$),
significantly lower than McGaugh. This difference must be mainly due to the 
higher surface brightness cut-off we find in our surface brightness
distribution compared to McGaugh, because the slopes at the faint end
of the distributions are similar.

\subsection{Semi-analytic Models}
\label{sams}

\ifthenelse{\value{emulate}=1}{
\begin{figure}[tb]
\epsfxsize=\linewidth \epsfbox[20 145 485 700]{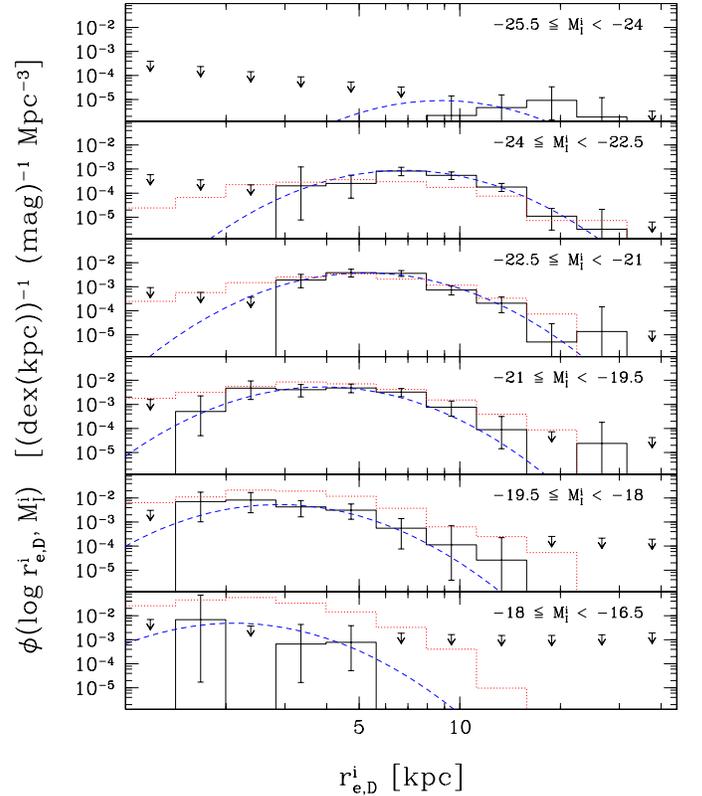}
\figcaption[f10.ps]{Bivariate space density distribution of Sb-Sdm galaxies as
a function of effective disk radius in different bins of total absolute
$I$-band magnitude as indicated in the
top-right corner of each panel. Symbols are the same as in
Fig.\,\ref{bivar_magtot_re}. The solid histogram and upper limits are
for the observed distribution, and the dashed line is the analytic fit
of equation\,(\ref{bivareq}). The dotted histogram is the prediction
of the 
semi-analytic galaxy formation model as described in the text.
\label{moddis}
}
\end{figure}
}{}

We will now compare our observed bivariate distribution with
theoretical predictions from the semi-analytic galaxy formation models
of Cole et al.~(\citeyear{Col00}). These models are based on the same
general scheme of galactic disk formation as was described in
\S\ref{funcdev}, but include much more physics in modeling the
evolution of both the dark matter and baryons, and relax some of the
simplifying assumptions made there, such as isothermal halos and
constant ratio of disk to halo mass. Here we simply
summarize the main ingredients of the models, and refer the reader to
Cole et al.~(\citeyear{Col00}) for a full description.

The starting point in the models is the initial spectrum of density
fluctuations.  The mass function of DM halos at any redshift is
calculated from this using the Press-Schechter~(\citeyear{PreSch74})
model. The formation of each halo through merging of smaller halos is
described by a merger tree. Merger trees are generated using a Monte
Carlo method also based on the Press-Schechter model. The process of
galaxy formation is followed through each halo merger tree. Gas
falling into halos is assumed to be shock-heated, and then to cool out
to the radius where the local radiative cooling time equals the halo
lifetime. The gas which cools collapses to form a rotationally
supported disk. Stars form from gas in the disk, on a timescale
related to the disk dynamical time. Supernovae are assumed to reheat
some of the gas and blow it out of the galaxy, with an efficiency
which is larger in small galaxies. Galaxies can merge, on a timescale
controlled by dynamical friction within halos, producing spheroidal
galaxies from disks. The chemical enrichment history of each galaxy is
calculated, and this is combined with the star formation history to
calculate the luminosity and colors of each galaxy using a stellar
population synthesis model. Finally, the effects of extinction by dust
are included.

Thus, in these models, the total baryonic mass of a galaxy is
determined by the combined effects of gas cooling from the halo, gas
ejection by supernova feedback, and mergers with other galaxies. The
result is a galaxy luminosity (and mass) function that has a
significantly different shape from the Press-Schechter mass function
of halos, and is close to the observed galaxy luminosity function.

The calculation of disk sizes proceeds along similar lines to those in
\S\ref{funcdev}. Each DM halo is assigned a total spin parameter
randomly drawn from the distribution (\ref{spineq}). The specific
angular momentum of the gas which cools is assumed to equal that of
the dark matter within the cooling radius, and gas is assumed to
conserve its angular momentum during the collapse down to a
centrifugally supported disk. As already discussed in \S\ref{funcdev},
this assumption of angular momentum conservation is valid in CDM-like
cosmologies only if feedback effects are strong enough to prevent most
of the gas condensing into dense lumps early on.  This is what is
assumed in the semi-analytic models, but most N-body/gasdynamics
simulations find that the cooling gas loses substantial angular
momentum, through forming dense lumps which then lose orbital angular
momentum to the DM halo by dynamical friction before merging together
to form galaxies. The resulting disks in these simulations are too
small compared to observed galaxies, and this currently constitutes
one of the most fundamental problems for galaxy formation models in
CDM-like cosmologies (e.g.\ Navarro \& Steinmetz \citeyear{NavSte00}).

Nonetheless, the semi-analytic models improve over the treatment in
\S\ref{funcdev} by having a physical calculation of the galaxy mass
and luminosity, by using Navarro, Frenk \& White (\citeyear{Nav97})
density profiles for the DM halos (which according to N-body
simulations is more appropriate in CDM-like universes than isothermal
spheres) and by including the self-gravity of the galaxy and the
contraction of the halo in response to the gravitational pull of the
galaxy. Thus, the disk radius is found by solving for the
self-consistent dynamical equilibrium of the disk, spheroid and DM
halo.

In Fig.\,\ref{moddis} we compare our bivariate luminosity-scalesize
distribution with the ``reference model'' of Cole et
al.~(\citeyear{Col00}), which is based on a CDM
cosmology with $\Omega_0=0.3$ and $\Lambda_0=0.7$.  The model assumed
$\sigma_{\lambda}=0.53$, based on N-body simulations. For the model we
only plot galaxies with bulge-to-total-light-ratio $<0.33$, equivalent
to Hubble types later than Sab. This includes Hubble types later than
Sd, which are not present in our observed sample. The model therefore
over-predicts the number of galaxies, especially at faint
luminosities, as late-type galaxies have a steeper faint end of the LF
as detailed in \S\,\ref{1Dproj}. The models do not provide any
detailed morphological information, only bulge-to-disk ratios, so we
have no means to remove the very late-type galaxy contribution from
the models.

At a given luminosity, the model predicts a scalesize distribution
that is somewhat broader than observed, especially at lower
luminosities.  This is the same discrepancy as we found in
\S\ref{func}, where the scalesize distribution for disks in
the simple-minded parameterization corresponded to
$\sigma_{\lambda}=0.36$. In fact, if the value of $\sigma_{\lambda}$
used in the semi-analytic model is reduced from 0.53 to 0.35, it also
gives a scalesize distribution with a very similar width to the
observed one. However, a value of $\sigma_{\lambda}$ this low does not
seem compatible with the results of N-body simulations of CDM-like
universes.

\subsection{The width of the disk size distribution: a conflict with
theory?}

We have seen that both the simple disk formation model described in
\S\ref{funcdev} and the more sophisticated semi-analytic models
described above result in a similar discrepancy with observations: the
width of the disk scalesize distribution at a fixed luminosity,
$\sigma(\ln\re)$, is predicted to be about 1.5 times larger than is
observed, if we use the value of $\sigma_{\lambda}$ from N-body
simulations, or, equivalently, that we need to assume a value of
$\sigma_{\lambda}$ about 0.7 times smaller than that measured in the
simulations in order to fit the observed scalesize distribution. How
might we explain the narrowness of the scalesize distribution within
the picture of hierarchical galaxy formation in a CDM-like universe?

The possibility that the true dispersion in halo spin parameters is
smaller than the value $\sigma_{\lambda}\approx 0.5-0.6$ that we have
assumed seems quite unlikely in a standard CDM-like universe. In
N-body simulations, the distribution of halo spin parameters is found
to be remarkably similar in different cosmologies, in different
density environments and for a large range of halo masses (e.g.\ Cole
\& Lacey \citeyear{ColLac96}, Lemson \& Kauffmann
\citeyear{LemKau99}). Even for self-interacting CDM (Spergel \&
Steinhardt \citeyear{SpeSte99}), this result would probably not change
much, as a halo acquires most of its angular momentum around
turnaround, when there is no significant difference in the behavior
of collisional and collisionless DM.

The specific angular momentum of the baryons that cool and collapse to
form the disk may be different from that of the dark halo as a whole,
but as long as the ratio of these two does not depend on the halo spin
parameter, the {\em fractional} width of the size distribution is
unaffected. For instance, in the semi-analytic models of Cole et
al.~(\citeyear{Col00}), the specific angular momentum of the gas which
cools is equal to that of the DM within the gas cooling radius, and so
is less than that of the halo as a whole, but scales with it, since
the cooling radius does not depend on the halo angular momentum.  Even
in a more complex model for the cooling of halo gas, which relaxes the
assumptions of a smooth spherical gas distribution, there is no
obvious reason why the relative width of the distribution of 
angular momentum of the gas that cools should be any different from
that of the halos, since
the rotation within the halo is dynamically unimportant and should not
affect which gas cools and collapses.
As already mentioned in \S\ref{sams}, N-body/gas-dynamics simulations
typically find that the gas loses angular momentum during merging and
collapse. This creates a difference between the disk and halo angular momentum,
but they are still correlated, albeit with substantial scatter (Navarro \& 
Steinmetz \citeyear{NavSte00}). This scatter will broaden the
predicted scalesize distribution, making the problem even worse.

Therefore, to reduce the width of the scalesize distribution, it seems
necessary to consider processes operating within galaxy disks after
they form. What is needed are processes which remove galaxy disks from
either the low or high angular momentum tails of the distribution. One
such process for removing low angular momentum disks was already
proposed by several authors (de Jong~\citeyear{Thes}; Dalcanton et
al.~\citeyear{Dal97b}; Mo, Mao \& White \citeyear{Mo98}; McGaugh \& de
Blok \citeyear{McGdeB98}). They noted that for a given disk mass, the
low angular momentum disks will be more strongly self-gravitating, and
so more likely to undergo bar instabilities, and suggested that disks
undergoing such instabilities would turn into spheroids, thus removing
disks of small sizes from the distribution. This would result in a
substantial population of spheroidal systems at all masses that were
not created by merging. An interesting test of this idea is whether it
predicts the correct luminosity and angular momentum distributions for
spheroids. Low luminosity elliptical galaxies are observed to have
significant rotation velocities, perhaps to the extent that they
cannot be explained by formation in major mergers (e.g.\ Rix, Carollo
\& Freeman \citeyear{Rix99}).

It is possible that the effects of star formation and/or feedback from
supernovae may suppress the number of large scalesize disks at a
given luminosity. There is observational evidence that the timescale
for star formation is shorter where the surface density of gas and
stars is larger (e.g.\ Schmidt \citeyear{Sch59}; Kennicutt
\citeyear{Ken89}; Dopita \& Ryder \citeyear{DopRyd94}). This naturally
gives rise to inside-out disk formation, as observations suggest for
most disk galaxies (Bell \& de Jong \citeyear{BeldeJ00a}). In addition,
it may be easier for supernova feedback to eject gas from the disk
where the surface density is lower, or from larger disk radii where
the escape velocity is lower. Both of these processes, star formation
and feedback, could therefore have the effect of reducing the total
luminosity of larger scalesize disks, for a given initial disk mass,
which might make the size distribution narrower at a given
luminosity. These processes could also result in disks where the scale
size of the stars is less than that of the gas which originally fell
in. If this effect is stronger in the larger scalesize, lower surface
density disks, this would also narrow the size distribution. The
semi-analytic models that we considered do not calculate the radial
dependence of star formation within a disk, but simply assume that the
scalesize of the stars and the gas are the same. The models also do
not include any explicit dependence of star formation or feedback on
surface density.

One final solution might be that the observational sample is biased
and that we suffer from morphological selection effects in our
comparison with the models. It could be that the largest and/or
smallest scalesize galaxies at each luminosity have preferentially
been classified as type later than Sdm. However, it is rather hard to
conceive how this could happen, as in many studies it has been found
that morphological type mainly correlates with luminosity and surface
brightness but is rather independent of scalesize (e.g.\ de Jong
\citeyear{deJ3}). The real test of this possibility awaits the proper
determination of the bivariate distribution of these very late type
galaxies.

In conclusion, we see that allowance for disk instabilities converting
disks into spheroids, more detailed physical calculations of star
formation and feedback in disks and/or morphological selection
effects, may well be able to explain the observed width of the disk
size distribution within the standard framework of hierarchical galaxy
formation.

\section{CONCLUSIONS}
\label{concl}

We have derived the bivariate space density distributions of Sb-Sdm
galaxies in luminosity, scalesize and surface brightness from
observational data by using a \vmax\ technique to correct for
selection biases.  A bivariate function described by
equation\,(\ref{bivareq}) was fitted to the observed distribution
using a maximum likelihood technique, and was found to fit the data
well. The main conclusions of this paper are as follows:

-- The bivariate space density distribution of spiral galaxies in the
(\magtot,\re)-plane is well described by a Schechter function in the
luminosity dimension and a log-normal scalesize distribution at a
given luminosity. The median disk size scales with luminosity as $\sim
L^{0.2}$ -- $L^{0.3}$.

-- This parameterization of the bivariate distribution was motivated
by a simple model for galaxy formation through hierarchical
clustering, where galaxies form in DM halos, which acquire their
angular momenta from tidal torques.  The galaxy luminosity
distribution is related to the distribution of halo masses, while the
disk scalesize distribution is related to the distribution of halo
angular momenta. However, although this model predicts the correct
{\em shape} for the disk size distribution, the {\em fractional width}
of this distribution is smaller than expected. The detailed
semi-analytic galaxy formation models of Cole et
al.~(\citeyear{Col00}) show a similar shortcoming. To make these
models consistent with the observations would require that
either the intrinsic angular momentum distribution of halos is
narrower than measured from N-body simulations, or additional physics
not included in the semi-analytic models is needed to describe the
formation of disks in DM halos.

-- The determination of the local LF of spiral galaxies is not
strongly effected by the bias against low surface brightness (LSB)
galaxies, even when selecting galaxies from photographic plates. This
may not be true for the deepest high redshift observations
available at the moment (the Hubble Deep Fields), where \zq\ surface
brightness dimming does cause a significant selection bias against LSB
galaxies at high redshifts.

-- The distribution of central surface brightnesses of spiral galaxy
disks integrated over all luminosities has a faint end slope similar to
the faint end slope of the LF.  This means that the number of spiral galaxies
per \magarc\ in a volume stays nearly constant when going to fainter
surface brightnesses, to the limit where we have been able to detect
galaxies (about 4 magnitudes below the canonical Freeman
(\citeyear{Fre70}) value of 21.65 $B$-\magarc).

-- The luminosity density of disk galaxies in the local universe is
dominated by fairly high surface brightness galaxies.  The contribution
of LSB galaxies to the local luminosity density is small, unless the galaxy
LF turns up dramatically at the faint end due to dwarf and irregular
galaxies.


\acknowledgements

We gratefully acknowledge Vince Ford, who provided the luminosity
profiles for all MFB galaxies in electronically readable format. We
thank the referee Stacy McGaugh for a constructive report.  Support
for R.S.~de Jong was provided by NASA through Hubble Fellowship grant
\#HF-01106.01-A from the Space Telescope Science Institute, which is
operated by the Association of Universities for Research in Astronomy,
Inc., under NASA contract NAS5-26555.  CGL acknowledges the support of
the Danish National Research Foundation through its establishment of
the Theoretical Astrophysics Center, and a PPARC Visiting Fellowship.
This work was partially supported by the PPARC rolling grant for
extragalactic astronomy and cosmology at Durham, by the EC TMR Network
on ``Galaxy Formation and Evolution'', and by a grant from ASI.  This
research has made use of NASA's Astrophysics Data System Abstract
Service, and of the NASA/IPAC Extragalactic Database (NED) which is
operated by the Jet Propulsion Laboratory, California Institute of
Technology, under contract with the National Aeronautics and Space
Administration.

\end{document}